\documentclass[a4paper,11pt]{article}
\usepackage[dvips]{graphicx} 
\usepackage[]{amsmath} 

\begin{document}

\vspace{50cm}

\begin{center}
\Huge{\textsc{Low-Energy Deuteron Polarimeter}}
\end{center}

\begin{figure}[!h]
\vskip5cm
\begin{center}
\scalebox{0.25}[0.25]{\includegraphics[]{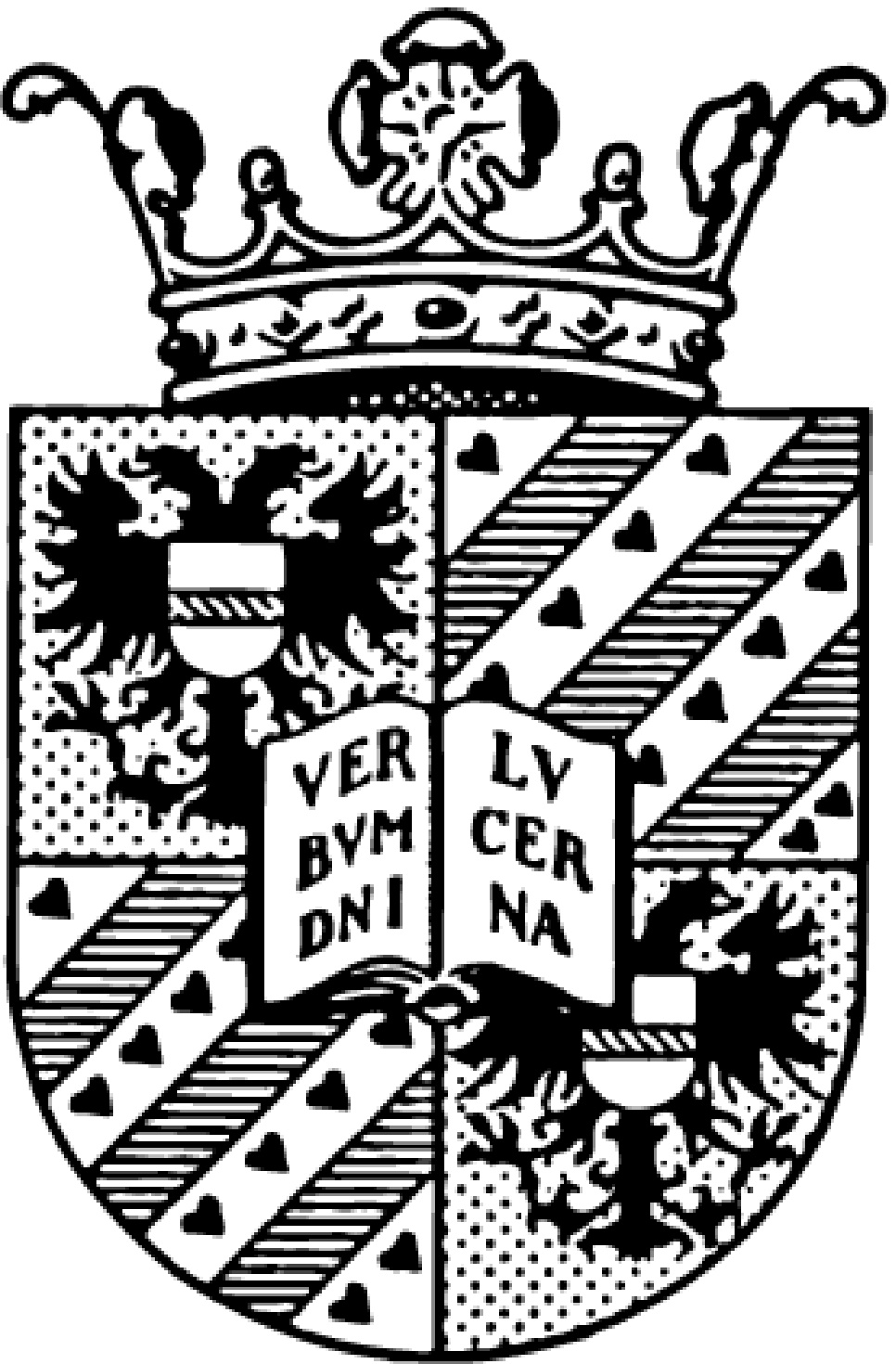}}
\end{center}
\end{figure}

\begin{center}
\vskip2cm
    \huge{Michael Bellos}\\
\end{center}

\vspace{0.5cm}
\begin{center}
        A thesis submitted\\
        to the University of Groningen\\
        in partial fulfillment of requirements\\
        for the degree of Masters of Science in Physics\\
        \bigskip
        \Large{Supervisors: Johan Messchendorp \\ 
           Nasser Kalantar-Nayestanaki}\\ 
        \bigskip
        January 2003--January 2004 \\
    \end{center}

\newpage

\begin{center}
\Large{\textbf{Abstract}}
\end{center}

\begin{large}
We built a Low Energy Deuteron Polarimeter (LDP) which measures the
spin-polarization of deuteron beams in the energy range of 25 to 80
keV. The LDP works by measuring \emph{azimuthal asymmetries} in the
\mbox{D($\vec d, n$)$^{3}$He} reaction at $\theta_{lab}=90^\circ$
and comparing them to \emph{analyzing powers}. We built this
polarimeter for two reasons. Firstly, to cross-check other
polarimeters at KVI (LSP and IBP). Secondly to test the LDP itself.
We were able to use the LDP as a vector polarimeter, but not as a
tensor polarimeter because of an uncertain tensor analyzing power
calibration.

\end{large}

\vfill
\begin{figure}[!h]\label{fig:asterix}
\hskip2cm {\includegraphics[]{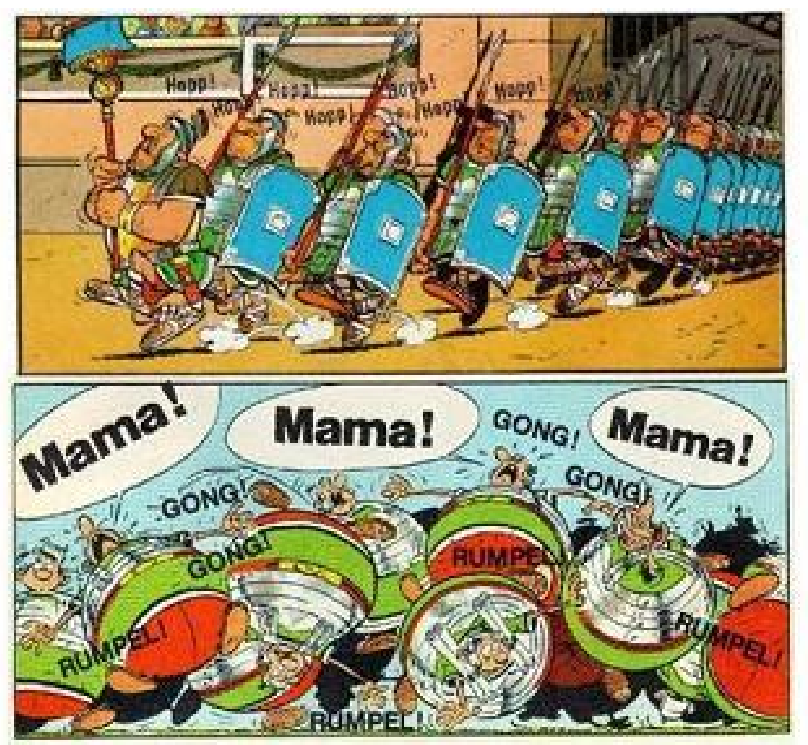}} \caption{\small{Cartoon
depicting polarization. The top panel represents a polarized beam,
where spins are aligned. The bottom panel represents an unpolarized
beam, where spins are randomly aligned.}}
\end{figure}

\begin{flushright}
Ver. 1.2\\
\end{flushright}

\newpage
\tableofcontents

\newpage
\section{Introduction}

Most people know that light can be polarized. This means that light
oscillates in a preferred plane. All nuclear physicists know that
`particle beams' can be `spin-polarized'. This means that particles
making up a beam have their spins aligned in a
preferred direction as is illustrated in Fig.\,1 (page 2).\\

The linear polarization of light can be determined, quite easily,
with a polarizing filter. An analogous device, called the
Stern-Gerlach filter, can determine the polarization of beams of
neutral particles by splitting the beam according to its spin
content. Here the degree of polarization would be the excess in
quantity of one spin-state
over another spin-state.\\

Most beams in nuclear physics are made of charged particles such as
electrons, protons, nuclei, or as in this experiment, deuterons. A
deuteron is an atomic nucleus consisting of a proton and a neutron.
Only beams of charged particles can be manipulated (accelerated,
steered, focused…) by the electric and magnetic fields of the
machinery in accelerator facilities, such as in
KVI\footnote{`Kernfysisch Versneller Instituut', The Dutch Nuclear
Physics Accelerator Institute}. Neutral beams are weakly affected by
electric and magnetic fields. Measuring the polarization of beams of
charged particles, as opposed to beams of neutral particles, is more
difficult because polarization filters such as the polaroid or the
Stern-Gerlach filters do not exist for charged particles. Mott and
Pauli ~\cite{mott,pauli}, therefore many textbooks in Quantum
Mechanics, claim that a spin filter for charged particles is
theoretically impossible. Yet it was recently claimed
\cite{batelaan} that such a device is possible under particular
conditions. Until such a device is built, or proved unfeasible, the
polarization of charged particle beams is determined by scattering
experiments. Measuring polarization may seem to be a trivial
measurement. But since it is a fully-fledged scattering experiment
it involves beam, target, detectors, and
electronics, therefore takes months to accomplish.\\

\subsection{Nomenclature}
Particle beam polarimetry borrowed nomenclature from optical
polarimetry. This is not surprising in view of the similarities
between spin polarization and optical polarization.\\

Physics textbooks often describe an experiment with two polarizing
filters at an angle to each other; the first filter is called the
`polarizer' and the second is called `analyzer'. The wording is such
because the first filter can polarize (normally unpolarized) light,
while the second filter can analyze the strength and direction of
the polarization. In nuclear physics polarimetry, a reaction of the
type\footnote{The notation A($b,c$)D represents a reaction where A
is the target, $b$ the beam (projectile), $c$ and D the observed
(ejectile) and unobserved (recoil) products, respectively. This
notation specifies more than $A+b \rightarrow c + D$ can.} A($b,
\vec c$)D is called a `polarization experiment'. The vector stands
for a polarized specie. In this reaction an unpolarized beam $b$,
creates a polarized ejectile $\vec c$. A reaction of the type
A($\vec b, c$)D is called `analyzing power experiment' because a
beam, $\vec b$, with non-zero polarization creates an asymmetry in
the ejectile's distribution. The word `analyzer' of optics became
'analyzing power' in nuclear physics. In optics any analyzer is as
good as the next one, while in nuclear physics different reactions
have different analyzing efficiencies, hence the word analyzing
\emph{power}.

\subsection{Why Make Polarized Beams?}

Polarized beams or targets are used in nuclear physics to extract
observables such as analyzing powers and spin-transfer coefficients.
These measured observables can be compared to predictions of
theoretical models to study, for example, the
three-body force \cite{3NF, ermisch}, or the spin-terms of the nucleon-nucleon potential.\\

One application of polarized beams lies in the possibility
\cite{kulsrud,Zhang} of using polarized deuteron beams to control
the reaction rates in fusion reactors and to reduce the amount of
unwanted neutrons.

\subsection{Our Motivation}
Our motivation to measure polarization stems from the fact that two
polarimeters at different beam energies measure different
polarizations. One polarimeter, called the In-Beam Polarimeter
(IBP), measures less polarization than another polarimeter, the
Lamb-Shift Polarimeter (LSP). By building a third polarimeter, this
Low-Energy Deuteron Polarimeter (LDP), and comparing its results
with the other two polarimeters, we wish to determine whether one of
the polarimeter is inaccurate or whether the polarization changes
between polarimeters. For beams of polarized protons, the LSP
routinely measures polarization between 80-90 $\pm$ 1\% (of the
theoretical maximum), while the IBP measures 70-75 $\pm$ 4\%. Since
laboratories, other than KVI, are not able to produce polarizations
reaching 90\%, a healthy skepticism exists in the LSP's reading of
90\%. This new polarimeter, the LDP, which measures essentially the
same beam as the LSP (in terms of beam energy, current and
location), can determine whether the LSP systematically
overestimates the polarization or not. If the LSP and LDP agree on
the polarization, one can conclude that there are polarization
transformations. While if the LDP agrees with the IBP, we conclude
that the LSP overestimates polarization due to
some unknown systematic uncertainty.\\

Another motivation is that, a deuteron polarimeter based on the
\mbox{D($d, n$)$^{3}$He} reaction has not been published before.
Building the LDP is a feasibility test for a new polarimeter. The
standard reactions, for deuteron polarimetry at these energies, are
$^{3}$H($d, n$)$^{4}$He and D($d, p$)$^{3}$H. The advantage of our
reaction over these two is, respectively, that it does not involve
handling of tritium, and does need to have detectors inside the
scattering chamber.

\subsection{Concept Behind Polarimetry}
\label{sec:polarimetry}

By impinging a beam on a target and observing the distribution of
scattered particles in space, one can deduce the polarization of the beam.\\

\begin{figure}[!h]
\hskip2cm \scalebox{0.6}[0.6]{\includegraphics[]{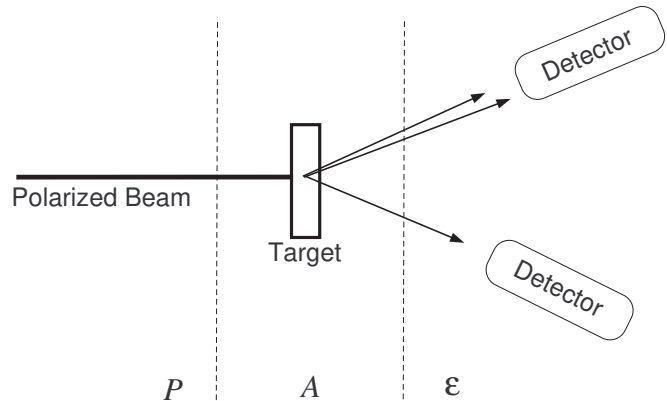}}
\caption{\small{A scattering experiment. A polarized beam impinges
on a target. The ejectiles scatter preferentially in some direction.
Two detectors measure an asymmetry in the number of ejectiles. The
figure also shows to which parts of the experiment the quantities
$P$ (polarization), $A$ (analyzing power), and $\epsilon$
(asymmetry) are associated with.}}\label{fig:scattering}
\end{figure}

An unpolarized beam scatters particles with an isotropic (azimuthal)
distribution, while a polarized beam scatters particles with a
\emph{non}-isotropic (azimuthal) distribution.
The latter case is illustrated in Fig.\,\ref{fig:scattering}.\\

The degree of non-isotropy in scattering is quantified by the
`asymmetry' that detectors measure, and is proportional to the beam
polarization with a proportionality constant given by the analyzing
power. Mathematically,

\begin{equation*}
\epsilon= P A.
\end{equation*}

This equation describes how polarimeters based on scattering, such
as the LDP and IBP, work. In our experiment $P$ is the quantity
which we solve for in terms of $\epsilon$ and $A$. $\epsilon$ is
measured experimentally by the detectors. $A$ is a constant that can
be calculated or measured by other experiments.

\newpage
\section{Theoretical Background}

\subsection{Nuclear Potential}
The nuclear potential is not yet fully understood, and is still a
subject of research and lamentation. An illustrative, but
incomplete, form of a two-nucleon nuclear potential is

\begin{multline}\label{eq:potential}
V=V_0(r)+V_s(r)\,\vec{s_1}\cdot\vec{s_2}+V_{SL}(r)\,\vec S \cdot
\vec L\\+V_T(r)\,\ \Big(\frac{3}{r^2}(\vec {s_1} \cdot \vec r)(\vec
{s_2} \cdot \vec r)-\vec {s_1}\cdot\vec{s_2}\Big) +
\underbrace{\ldots}_{?}
\end{multline}\\

Terms on the right side of the above equation are called central,
spin-spin, spin-orbit, and tensor. The central term depends only on
the distance separating the two nucleons. Another example of a
central potential is gravitational attraction. The spin-spin term,
$\vec{s_1}\cdot\vec{s_2}$, stems from the magnetic interaction of
the spins of the two nucleons. The spin-orbit term, $\vec S\cdot
\vec L$, stems from the interaction between the total spin of both
nucleons and the angular momentum defined by their relative motion.
An example of a spin-orbit interaction is the `fine structure' in
atomic physics; where the degeneracy for states of equal $L$ is
lifted. An analogy of a tensor behavior is the interaction of two
bar magnets. Two bar magnets with parallel orientations placed
alongside each other (like sardines in a can) will repel each other.
Two bar magnets with parallel orientation placed along a line (like
sardines chasing each other) will attract. The tensor term of the
nuclear potential exhibits the same angle-dependent behavior.

\subsection{Spin-Orbit Term}
The spin-orbit term of the nuclear potential is responsible for the
azimuthal distributions of scattered particles. Spin-spin and tensor
terms contribute to other polarization phenomena not addressed is
this experiment. In case $V_{SL}(r)$ or $\vec S\cdot \vec L$ is zero
in Eq.\,(\ref{eq:potential}), one would observe a flat distribution
of ejectiles in the azimuthal angle.\\

\begin{figure}
\hspace{1.7cm}
\includegraphics[width=0.5\textheight]{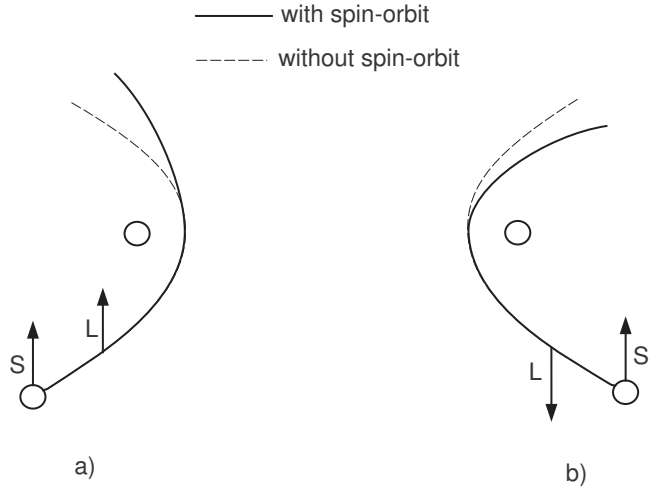}
\caption{\small{Scattering of a particle with spin $S$ on a spinless
target. The collision defines the angular momentum $L$. In a) $S$
and $L$ parallel, reducing the attractive nuclear force. In b) $S$
and $L$ are antiparallel, increasing the attractive nuclear force.
Without the spin-orbit term trajectories a) and b) would be
indistinguishable.}} \label{fig:ls}
\end{figure}

Assume, without loss of generality, that $V_{SL}(r)>0$. When $\vec
S$ and $\vec L$ are parallel, the spin-orbit term is positive, and
therefore decreases the attractive negative nuclear potential. This
is illustrated in Fig.\,\ref{fig:ls}\,a, where the trajectory of a
particle in a collision gets less deviated in the presence of
parallel spin and orbit angular momentum. Conversely, if $\vec S$
and $\vec L$ are antiparallel, then the potential will be more
attractive, and the particle will scatter more towards the target
nucleus as illustrated in Fig.\,\ref{fig:ls}\,b. In both cases
particles are scattered preferentially towards the right
direction.\\

\subsection{Polarization Formalism}
Every deuteron has spin. Furthermore, this spin must be aligned in
one of three ways with respect to a quantization axes. The spin can
be aligned parallel ($m_I=1$), anti-parallel ($m_I=-1$), or
perpendicular ($m_I=0$) to this quantization axis. \\

The polarization formalism given below is based on articles by
Ohlsen \cite{ohlsen, ohlsen2} and describes the relation between all
quantities involved in this project. It describes how quantum
mechanics, experimental and theoretical nuclear physics meet. This
formalism allows one to derive few simple equations
(Eq.\,(\ref{eq:spin1/2asym}) \& Eq.\,(\ref{eq:spin1asym})) that are
used to determine beam polarization in terms of known
analyzing powers and measured experimental asymmetries.\\

Although deuterons are spin-1 particles, I will start the
polarization formalism for spin-$\frac{1}{2}$ particles then extend
it to spin-1 particles. The reason is that, the polarization
formalism for spin-$\frac{1}{2}$ particles contains all the
essential features of the formalism with much less mathematics.
Therefore it is a good starting point to visualize.

\subsubsection{Coordinate System}

\begin{figure}
\hskip1.5cm
\scalebox{0.9}[0.9]{\includegraphics[width=1\textwidth]{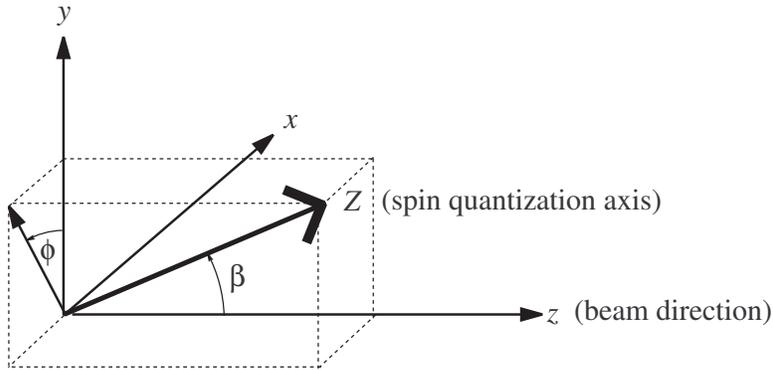}}
\caption{\small{Coordinate system used to describe scattering with a
polarized beam.}} \label{fig:coordsys}
\end{figure}

The coordinate system most often used to describe polarization
experiments is called the `Madison Convention' \cite{barschall} and
is shown in Fig.\,\ref{fig:coordsys}. This coordinate system
incorporates both beam and scattering parameters into one coordinate
system. The direction of the $z$-axis is parallel to the momentum of
the incoming beam, \emph{\textbf{k}}$_{in}$. The $y$-axis is along
\emph{\textbf{k}}$_{in}\times$ \emph{\textbf{k}}$_{out}$, where
\emph{\textbf{k}}$_{out}$ is the direction of the outgoing ejectile.
The $y$-axis, therefore, is perpendicular to the scattering plane.
The $x$-axis is left to form a right-handed system with the $y$ and
$z$ axes. The angle between $z$ and $Z$ is given by $\beta$. The
angle $\phi$ is between $y$ and the projection of $Z$ onto the
$x$-$y$ plane. Scattering to the the left, right, up and down with
respect to the quantization axis correspond to $\phi=$ 0, 180, 270
and \mbox{90$^\circ$} respectively.

Note that $z$ (lower-case) is the beam direction, while $Z$
(upper-case) is the quantization axis direction. One must
distinguish between $p_z$ and $P_Z$. The former is the component of
polarization along the the beam direction, while the latter is the
degree of polarization along the quantization axis, the quantity of
interest in this experiment.

\subsubsection{Spin-$\frac{1}{2}$ Particles}
A spin-$\frac{1}{2}$ particle can be represented by a Pauli
spinor,\\
\begin{equation}\label{eq:spin}\chi= \left(\begin{array}{c}
  a_1 \\
  a_2 \\
\end{array}\right)
=a_1 \left(\begin{array}{c}
  1 \\
  0 \\
\end{array}\right)
+ a_2 \left(\begin{array}{c}
  0 \\
  1 \\
\end{array}\right),
\end{equation}\\
where $a_1$ is the probability amplitude of finding the particle in
the state $\left(\begin{array}{c}1 \\0
\\ \end{array}\right)$ called spin-up, and $a_2$ is the probability
amplitude of finding the particle in the spin-down state
$\left(\begin{array}{c}0 \\1\\ \end{array}\right)$.\\

The spin state of an ensemble of $N$ such particles can be
represented
 by a set of Pauli spinors,\\*
\[\chi^{(n)}= \left(\begin{array}{c}
  a_1^{(n)} \\
  a_2^{(n)} \\
\end{array}\right) \qquad n=1,\ldots, N\,,\]\\
where $n$ runs through all particles.\\

If the order of spins in an ensemble is not important, but only the
average spin, then a beam can be described by a density
matrix,\\

\begin{equation}
\rho=\left(%
\begin{array}{cc}
  \sum_{n=1}^{N}|a_1^{(n)}|^2 & \sum_{n=1}^{N}a_1^{(n)}\,a_2^{(n)*}    \\
  \sum_{n=1}^{N}a_2^{(n)}\,a_1^{(n)*} & \sum_{n=1}^{N}|a_2^{(n)}|^2 \\
\end{array}%
\right).
\end{equation}\\

A density matrix fully characterizes the polarization (magnitude
and direction) of particle beams.\\

Here are some relations,

\begin{eqnarray} \label{}
\nonumber  p_x &=& Tr(\rho \, \sigma_x)= \frac{1}{N}\sum_{n=1}^N 2
\,
Re(a_1^{(n)} \, a_2^{(n)*}),\\
 p_y &=& Tr(\rho \, \sigma_y)= \frac{1}{N}\sum_{n=1}^N 2 \,Im(a_1^{(n)}
\, a_2^{(n)*}),\\
\nonumber p_z &=& Tr(\rho \, \sigma_x)= \frac{1}{N}\sum_{n=1}^N
(|a_1^{(n)}|^2 - |a_2^{(n)}|^2).
\end{eqnarray}

One can show \cite{ohlsen} that the density matrix can be written as
a linear combination of Pauli spin operators,

\begin{equation}
\rho=\frac{1}{2}\,(I+\sum_{j=1}^3 p_j \, \sigma_j),
\label{eq:ohlsen}
\end{equation}\\
where:
\begin{itemize}
\item   $p_j$\,'s are the components of polarization in the $x$, $
y$, and $z$ directions;
\item  $I$ is the unit matrix $\left(\begin{array}{cc}  1 & 0 \\
0
        & 1 \\ \end{array}\right);$
\item   $\sigma_j$\,'s are the Pauli spin operators;\\\\
        $\sigma_x=\left(\begin{array}{cc}  0 & 1 \\  1 &  0 \\
        \end{array}\right);$\quad$\sigma_y=\left(\begin{array}{cc}  0 & -i \\  i &  0 \\
        \end{array}\right);$\quad$\sigma_z=\left(\begin{array}{cc}  1 & 0 \\  0 &  -1 \\
        \end{array}\right).$\\
\end{itemize}

A nuclear reaction can transform the spin state of particles.
Therefore, the spinor of an outgoing particle $\chi_f$ is related to
the spinor of an incoming particle $\chi_i$  by a transformation,
$M$,

\[\chi_f=M\,\chi_i \,.\]\\
$M$ is a $2\times 2$ matrix whose elements are functions of energy
and
angle.\\

The density matrix describing the incoming beam $\rho_i$ can be
written in terms of spinors as

\[\rho_i=\sum_{n=1}^N \chi_i^{(n)}[\chi_i^{(n)}]^\dag ,\]
and for the outgoing beam,
\[\rho_f=\sum_{n=1}^N \chi_f^{(n)}[\chi_f^{(n)}]^\dag .\]
The density matrix is transformed by a reaction as

\begin{equation}\label{eq:rhotransf}
\rho_f=M\,\rho_i\,M^\dag .
\end{equation}\\

The cross section of the reaction can be given by,
\begin{equation}\label{eq:crosssec}
\sigma(\theta,\phi)=\frac{Tr(\rho_f)}{Tr(\rho_i)}
=\frac{Tr(M\,\rho_i\,M^\dag)}{Tr(\rho_i)}\,.
\end{equation}\\
If the beam is unpolarized

\[\rho_i\propto\left(%
\begin{array}{cc}
  1 & 0 \\
  0 & 1 \\
\end{array}%
\right),\]\\
and if the density matrix is normalized to unity

\[\rho_i=\frac{1}{2}\left(%
\begin{array}{cc}
  1 & 0 \\
  0 & 1 \\
\end{array}%
\right),\] \\
then Eq.\,(\ref{eq:crosssec}) reduces to

\[\sigma_0=\frac{1}{2} Tr(M\,M^\dag) \qquad \textrm{(unpolarized
beam)}.\]\\
Applying Eq.\,(\ref{eq:rhotransf}) to Eq.\,(\ref{eq:ohlsen}) one
gets,

\[\rho_f=\frac{1}{2} M\,M^\dag+\frac{1}{2} \sum_{j=1}^3
p_j\,M\sigma_jM^\dag.\] \\
Taking the trace yields

\begin{equation}
\label{most}
\sigma(\theta,\phi)=Tr(\rho_f)=\sigma_0\big(1+\sum_{j=1}^3
p_j\,A_j(\theta)\big),
\end{equation}\\
where

\[A_j=\frac{Tr(M\sigma_jM^\dag)}{Tr(MM^\dag)} \qquad j=x, y,
z.\]\\
$M$ is the same $2\times2$ matrix that was used to transform the
spin state in Eq.\,(\ref{eq:spin}) and density matrix in
Eq.\,(\ref{eq:rhotransf}). $A_j$'s are the analyzing powers of the
reaction. The analyzing powers, like cross section, are a property
of nuclear reactions. They can be calculated from theory or
measured experimentally.\\

Parity and time reversal arguments \cite{ohlsen} reduce
Eq.\,(\ref{most}) for two-body reactions to

\begin{equation}\label{eq:spin1/2y}
\sigma(\theta,\phi)=\sigma_0(1+ p_y\,A_y(\theta)).
\end{equation}\\
This implies that reactions are only sensitive to polarization along
the $y$-axis. Polarization along the $x$,
or $z$ direction do not affect the cross section. \\

\begin{figure}
\hskip1cm
\includegraphics[]{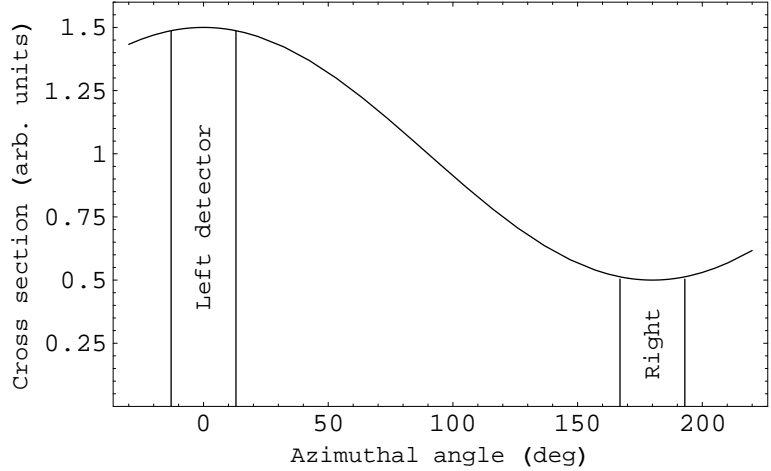}
\caption{\small{Azimuthal angle dependence of the cross section for
$P_Z=1$ and $A_y=0.5$\,. In this example there are more particles
going to the left direction than to the right direction. Detectors
placed in these directions measure an
asymmetry.}}\label{fig:csdance}
\end{figure}

Since $p_y=\vec{P_Z}\cdot\vec y=P_Z\cos\phi$, \,
Eq.\,(\ref{eq:spin1/2y}) can be written as

\begin{equation}
\label{eq:spin1/2PZ} \sigma(\theta,\phi)=\sigma_0\big(1+
P_Z\cos\phi\,A_y(\theta)\big),
\end{equation}\\
which is plotted in Fig.\,\ref{fig:csdance}\,.\\

Scattering to the left of the quantization axis ($\phi =0$) has a
cross section

\[\sigma_L=\sigma_0(1+P_Z\,A_y).\]\\
A detector placed in this direction will measure a count
proportional to this cross section
\[ L \propto \sigma_L .\]

Scattering to the right of the quantization axis ($\phi =180$) would
have a cross section of,

\[\sigma_R=\sigma_0(1-P_Z\,A_y).\]\\
A detector placed in this direction  will measure a count
proportional to
\[ R \propto \sigma_R .\]

Define the asymmetry, $\epsilon_1$, of the reaction as,

\[\epsilon_1\equiv\frac{L-R}{L+R}.\]\\
The asymmetry is, therefore, the difference over the sum of the
number of particles detected in the left and right
detectors.\\

Scattering in a direction parallel to the quantization axis has the
same cross section as scattering in an anti-parallel direction.
Therefore two detectors placed above and below the scattering will
measure the same number of ejectiles, resulting in
no asymmetry.\\

It can be seen that

\begin{equation}\label{eq:spin1/2asym}
P_Z\,Ay=\epsilon_1.
\end{equation}\\
This equation relates the polarization, analyzing power, and
asymmetry for a spin-$\frac{1}{2}$ beam. We introduced this equation
in Sec.\,\ref{sec:polarimetry}, and now present it with subscripts
reflecting some of the geometry behind the scattering.

\subsubsection{Spin-1 Particles}

The cross section of a reaction using a spin-1 polarized beam is

\begin{multline*}
\sigma(\theta,\phi)=\sigma_0\Big(1+ \frac{2}{3}p_yA_y(\theta)
+\frac{3}{2}p_{xz}A_{xz}(\theta)\\+\frac{1}{3}p_{xx}A_{xx}(\theta)
-\frac{1}{3}p_{yy}A_{yy}(\theta)+\frac{1}{3}p_{zz}A_{zz}(\theta)\Big).
\end{multline*}\\
This equation is the spin-1 analogy of Eq.\,(\ref{eq:spin1/2y}).
Yet it contains noticeably more terms!\\

By expressing the polarization in terms of the coordinates
($P_Z,P_{ZZ},\beta,\phi$) instead of
($p_y,p_{xz},p_{xx},p_{yy},p_{zz}$) we get,

\begin{multline*}
\sigma(\theta,\phi)=\sigma_0\Big(1+\frac{3}{2}P_ZA_y(\theta)\cos\phi
\sin\beta-P_{ZZ}A_{xz}(\theta)\sin\beta\cos\beta\sin\phi\\
-\frac{1}{4}P_{ZZ}\big(A_{xx}(\theta)-A_{yy}(\theta)\big)\sin^2\beta\cos2
\phi+\frac{1}{4}P_{ZZ}A_{zz}(\theta)\big(3\cos^2\beta-1\big)\Big).
\end{multline*}

In the spin-$\frac{1}{2}$ case, only one polarization ($P_Z$) and
one analyzing power ($A_y$) enter the equations, while for spin-1
beams, two polarizations ($P_Z$, $P_{ZZ}$) and four analyzing powers
($A_y$,  $A_{xz}$, $A_{xx}-A_{yy}$, $A_{zz}$) enter the equations.
This complexity arises because a spin-1 beam has three
spin substates, while a spin-$\frac{1}{2}$ beam only two (spin up and spin down).\\

It is worth mentioning that quantities with one index are called
vectors, while quantities with two indices are called tensors. For
example, $P_Z$ is called  `vector polarization', and $A_{zz}$ is
one of the three `tensor analyzing powers'.\\

Scattering to the left, right, up, and down ($\phi=$ 0$^\circ$,
180$^\circ$, 270$^\circ$, 90$^\circ$) directions with respect to the
quantization axis have cross sections,

\begin{eqnarray*}
  \sigma_L &=& \sigma_0(1+\frac{3}{2}P_Z A_y(\theta) \sin\beta+
        \frac{1}{2}P_{ZZ}(A_{yy}(\theta)\,\sin^2\beta +A_{zz}\,\cos^2\beta)), \\
  \sigma_R &=& \sigma_0(1-\frac{3}{2}P_Z A_y(\theta) \sin\beta+
        \frac{1}{2}P_{ZZ}(A_{yy}(\theta)\,\sin^2\beta +A_{zz}\,\cos^2\beta)), \\
  \sigma_U &=& \sigma_0(1+P_{ZZ} A_{xz}(\theta) \sin\beta\cos\beta+
        \frac{1}{2}P_{ZZ}(A_{xx}(\theta)\,\sin^2\beta +A_{zz}\,\cos^2\beta)), \\
  \sigma_D &=& \sigma_0(1+P_{ZZ} A_{xz}(\theta) \sin\beta+ \frac{1}{2}P_{ZZ}(A_{yy}(\theta)\,\sin^2\beta\cos\beta
        +A_{zz}\,\cos^2\beta)).
\end{eqnarray*}\\
Detectors placed in these scattering directions will measure a count
($L$, $R$, $U$, $D$) proportional to the cross sections,

\begin{eqnarray*}
  L &\propto& \sigma_L \\
  R &\propto&  \sigma_R\\
  U &\propto&  \sigma_U \\
  D &\propto& \sigma_D.
\end{eqnarray*}\\
One defines the five asymmetries $\epsilon_1$, $ \epsilon_2$,
$\epsilon_3$, $\epsilon_4$, $\epsilon_5$, by the following equations

\begin{eqnarray}\label{eq:spin1asym}
\nonumber  \epsilon_1 &\equiv& \frac{L-R}{L+R}=
\frac{\frac{3}{2}P_Z\sin\beta
    \, A_y}{1+\frac{1}{2}P_{ZZ}[\sin^2\beta A_{yy}+\cos^2\beta A_{zz}]}   \\
\nonumber  \epsilon_2 &\equiv&  \frac{U-D}{U+D}=
\frac{P_{ZZ}\sin\beta \,
  \cos\beta \, A_{xz}}{1+\frac{1}{2}P_{ZZ}[\sin^2\beta A_{xx}+\cos^2\beta A_{zz}]}\\
  \epsilon_3 &\equiv& \frac{2(L-R)}{L+R+U+D}= \frac{\frac{3}{2}P_Z\sin\beta
    \, A_y}{1+\frac{1}{4}P_{ZZ}[3(\cos^2\beta-1)A_{zz}]}  \\
\nonumber  \epsilon_4 &\equiv& \frac{2(U-D)}{L+R+U+D}=
\frac{P_{ZZ}\sin\beta \,
  \cos\beta \, A_{xz}}{1+\frac{1}{4}P_{ZZ}[3(\cos^2\beta-1)A_{zz}]}  \\
\nonumber  \epsilon_5 &\equiv& \frac{(L+R)-(U+D)}{L+R+U+D}=
\frac{-\frac{1}{4}P_{ZZ}\sin^2\beta \,
  \,
  (A_{xx}-A_{yy})}{1+\frac{1}{4}P_{ZZ}[3(\cos^2\beta-1)A_{zz}]},
\end{eqnarray}\\
which relate the asymmetries, analyzing powers, and polarizations
for spin-1 beams. These are the generalization from
spin-$\frac{1}{2}$ (Eq.\,(\ref{eq:spin1/2asym})) to spin-1
particles.

\subsection{$d+d$ Reactions}\label{sec:d+dreaction}

What happens when a deuteron beam strikes a deuteron target? Many
things happen, so let's restrict the question to: What nuclear
reactions are possible when a deuteron beam strikes a deuteron
target? Table \ref{tab:d+d} lists all known reactions.\\

\begin{table}
\hskip1cm
\begin{tabular}{l c c }
Reaction type & Reaction & Q-value (MeV)\\
\hline Elastic scattering\footnotemark[1] & D$(d, d)$D& 0\\
One-particle rearrangement      &D$(d, n)^3$He          & 3.27\\
                                 &D$(d, p)^3$H          &4.03\\
Radiative capture               &D$(d, \gamma)^4$He     &23.8\\
Breakup reaction\footnotemark[2]                &D$(d, n)p$\,D  &-2.2\\
\end{tabular}
\caption{\small{$d+d$ nuclear reactions.}}\label{tab:d+d}
\end{table}

\footnotetext[1]{In elastic
 scattering total kinetic energy is conserved. For example, billiard
  ball collisions. All particles in D($d,d$)D are deuterons. In this
   notation capital letters represent an atom or molecule, while
    lower-case letters represent a nucleus. D is called deuterium, while
     $d$ is called deuteron. These particles differ only by one electron.}
\footnotetext[2]{This is the only reaction that does not take place
at our beam energy. Beam energy needs to exceed the Q-value for the
reaction to occur, due to conservation of energy.}

A number of reaction listed in Table \ref{tab:d+d} take place in the
center of the sun, and possibly in future fusion reactors.
Coincidentally, solar and reactor plasmas are in the same energy
range as in this experiment. Yet this setup is used to
measure polarization, and not to study solar plasmas.\\

\begin{figure}
\hskip0.8cm \scalebox{1}[1]{\includegraphics[]{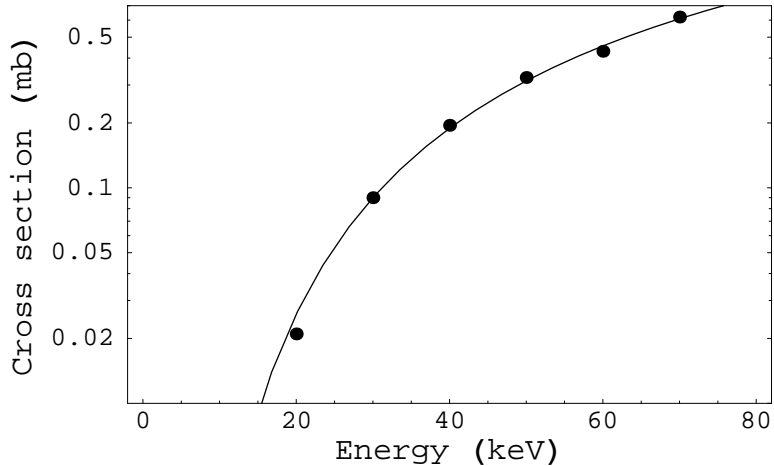}}
\caption{\small{Energy dependence of the D$(d,n)^3$He cross section.
Fit of a Gamov model to data points from various experiments.}
}\label{fig:crsd}
\end{figure}

The cross section of the D$(d,n)^3$He reaction highly depends on the
incident deuteron energy, as is shown in Fig.\,\ref{fig:crsd}. A
simple quantum mechanical effect can model this dependence
satisfactorily, as is explained below.

The potential between two nuclei is attractive at short distances (a
few fm) because of the strong force, and repulsive at large
distances (greater than a few fm) because of the coulomb force. A
deuteron approaching another deuteron sees a potential barrier with
a height of about 1 MeV. How can a deuteron with kinetic energy in
keV range get through the barrier to fuse with the other deuteron?
The answer is quantum mechanical tunnelling! The tunnelling
probability, hence the cross section, is proportional to

\begin{equation}
\sigma(E)\propto e^{\frac{-1}{\sqrt{E}}}
\end{equation}\\
This dependence is called the `Gamov factor'. Fitting it to
published data sets \cite{nndc} of D$(d,n)^3$He cross sections
yields an empirical relation,

\begin{equation}\label{eq:empcs}
\sigma(E)=a \, e^{\frac{-b}{\sqrt{E}}}
\end{equation}\\
with $a=22\pm 4$\, mb and $b=30\pm 2\, \sqrt{\textrm{keV}}$.

\subsection{Stopping Power}\label{sec:stoppingpower}

The basis of any understanding in experimental nuclear and particle
physics depends on the understanding of the `passage of radiation
though matter'. Instances of `passage of radiation through matter'
are; an alpha beam impinging on a gold foil (Rutherford's famous
experiment), a neutron depositing energy in a detector (as in this
experiment), and ion radiotherapy (where ion
beams destroy tumors).\\

As charged particles pass through matter they lose energy and are
deflected. This is not surprising in light of the many imaginable
ways in which particle and matter can interact. The most important
phenomena that contribute to the net process are

\begin{enumerate}
    \item Inelastic
collisions with atomic electrons
    \item Elastic scattering from nuclei
    \item Inelastic nuclear reactions
    \item Cherenkov radiation
    \item Bremsstrahlung
    \item etc \ldots
\end{enumerate}

A formula which models these phenomena is the well-known Bethe-Bloch
formula \cite{leo}. This semi-empirical formula gives the energy
loss ($\frac{dE}{dx}$) of various charged particles though various
materials. With the energy loss---also called stopping power---one
can calculate the range of particles and total energy deposited in
matter. These quantities are vital for detector and safety
consideration, since you want to know where your particles are going
and with how much energy! Although the Bethe-Bloch formula for
stopping power is extensively used in nuclear physics, it is only
valid for particles with energies greater than 1 MeV/nucleon. At the
low energies of this experiment another model and formula for
stopping power is given by Lindhard \cite{lindhard}, which states
that the stopping power is proportional to the beam velocity

\begin{equation}
\frac{dE}{dx} \propto v \propto \sqrt{E}.
\end{equation}

Anderson \cite{anderson} has compiled empirical formulas for
stopping powers of various beam and target combinations, by fitting
data with the Bethe-Bloch and Lindhard formulas. The stopping power
of a deuteron beam on a C$_2$D$_4$ target is given\footnote{Use
Bragg's Rule and the the assumption that the energy loss of
deuterons through deuterium is equal to that of protons through
hydrogen.} by

\begin{equation}\label{eq:empdedx}
\frac{dE}{dx}=14.48\,E^{0.45}\qquad\textrm{($[E]=$ keV and
$[x]=\mu$m).}
\end{equation}

Calculation of the mean penetration depth of the beam into the
target yields 0.4 to 0.7 $\mu$m for $E_d=$ 25 to 80 keV. This is the
distance within which the mean beam energy decreases to zero. Since
many relevant parameters---such as cross section and analyzing
power---are energy dependent, their average value has to be
calculated. The cross section is energy dependent, and since there
is an energy loss of the beam through the target, therefore
the average cross section must be calculated.\\

The Bethe-Bloch Formula does not apply to neutral particles such as
photons and neutrons. Neutral particles have a larger range through
matter than charged particles. Particles created from a $d+d$
reaction, at these energies, include neutrons ($n$), protons ($p$),
tritons ($^3$H), helium-3 ($^3$He), helium ($^4$He), and
$\gamma$-rays. $\gamma$-rays are also observed from the
de-excitation of nuclei after having absorbed neutrons. Therefore,
wherever one observes neutrons, one is likely to observe
$\gamma$-rays as well. The charged particles ($p$, $^3$H, $^3$He,
and $^4$He) do not make it out of the vacuum chamber. They are
stopped in the glass beam tube, metal target holder, or the target
itself due to their small range of order order $\mu$m to mm. This
tiny range is due to the high energy loss of low energy charged
particles passing through matter. Neutral particles, on the other
hand, such as neutrons and $\gamma$-rays, do not interact as much
with material and are able to exit the beam tube and reach the
detectors or go beyond them. This is why the D($d, n$)$^3$He
reaction was chosen, because it could be isolated from other
reactions also occurring.

\newpage
\section{Experimental Set-Up}

\begin{figure}[!h]
\hskip0.7cm \scalebox{0.6}[0.5]{\includegraphics[]{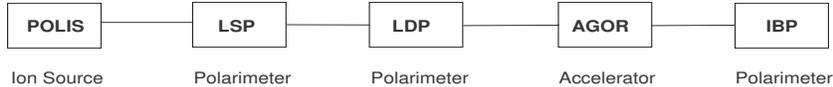}}
\caption{\small{Schematics of the beam line we used. Polarized beams
are created by POLIS, the beam polarization is measured with the LSP
and the LDP, the beam is accelerated, and finaly the beam
polarization is measured with the IBP.}} \label{fig:beamline}
\end{figure}

The different devices of our experiment are sketched in
Fig.\,\ref{fig:beamline}.\\

We briefly describe POLIS. We describe at length the components of
the LDP. The LSP \cite{kremerslsp, lsp2}, AGOR \cite{agor}, and the
IBP\cite{ibp} are described elsewhere.
\subsection{Polarized Ion Source}

Our Polarized Ion Source \cite{kremerspolis} (POLIS) provides beams
of polarized protons or deuterons. Proton beams can be vector
polarized, while deuteron beams can be vector and/or tensor
polarized.\\

Before the invention of polarized ion sources, such as POLIS,
polarized beams were produced by using the scattered ejectiles of a
reaction. These ejectiles were partly polarized, and were used
themselves as a beam for another experiment. These `double
scattering' experiments were plagued by low polarization and
intensity.\\

The production of a polarized deuteron beam from POLIS is outlined
as follows,

\begin{center}
$\rm{D}_2 \xrightarrow{dissociation} D \xrightarrow{transitions}
D^*\xrightarrow{ionization}$ $\overrightarrow d$
\end{center}

First, deuterium molecules from a gas cylinder are dissociated into
deuterium atoms and collimated into a beam. Then, electromagnetic
transitions polarize the atom by populating some of its hyperfine
states. The polarized atoms are ionized leaving only a beam of
polarized deuteron nuclei. This beam is then accelerated
to desired energies, and steered to a particular experiment.\\

\begin{figure}[!h]
\hskip1.5cm
\scalebox{0.8}[0.8]{\includegraphics[]{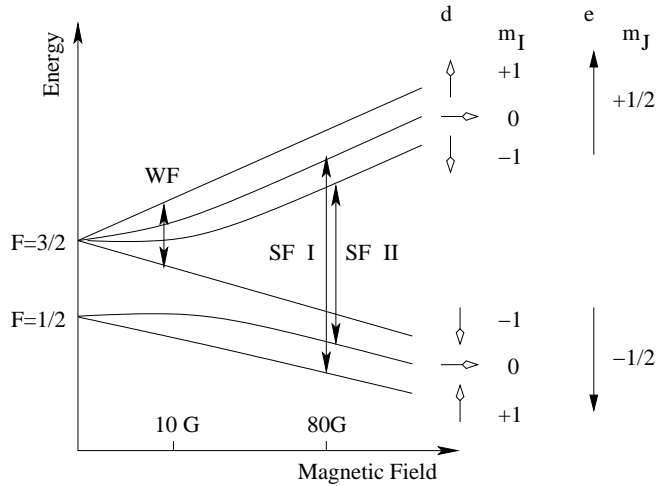}}
\caption{\small{Breit-Rabi diagram of a deuterium atom in the ground
state. The spin substate of the nucleus (d) and its electron (e) are
listed on the right of the figure.}} \label{fig:deuteriumlevels}
\end{figure}

Each hyperfine state of an atom represents a particular alignment of
the spin of the nucleus and electrons. The electromagnetic
transitions between hyperfine states are produced by applying RF
fields to atoms present in a magnetic field, as depicted in
Fig.\,\ref{fig:deuteriumlevels}. POLIS has four transitions units
called Weak Field, Medium Field, Strong Field I, and Strong Field
II. Different combination of these transition units populates
different hyperfine states. Populating a particular hyperfine states
is selecting a spin substate of the nucleus and electrons. The spin
substate of electrons is irrelevant since electrons are stripped
from the atom. The spin substate of the nuclei constitute the beam
polarization.

\begin{table}[]
\hskip 1cm
\begin{small}
\begin{tabular}{|c||c|c|c|}
 \hline
 POLIS &\multicolumn{2}{c|}{Max. Theoretical}&Beam \\
 Transition &\multicolumn{2}{c|}{Polarization}& Polarization\\
\cline{2-3}
Units&$P_Z$&$P_{ZZ}$&\\
\hline\hline
 WF  &$-2/3\quad$&0&Positive Vector\\
 SF. I + SF. II&$2/3$&0&Negative Vector \\
 MF + sextupole + SF. I&0&1&Positive Tensor \\
 MF + sextupole + SF. II&0&-2&Negative Tensor\\
\hline
\end{tabular}
\end{small}
\caption{\small{POLIS transition units with the beam polarization
they generate. Other combination of transition units are possible
yielding beams with both non-zero $P_Z$ and
$P_{ZZ}$.}}\label{tab:polis}
\end{table}

\subsection{Low Energy Deuteron Polarimeter}

\begin{figure}[!h]
\hskip 0.5cm \scalebox{0.8}[0.8]{\includegraphics[]{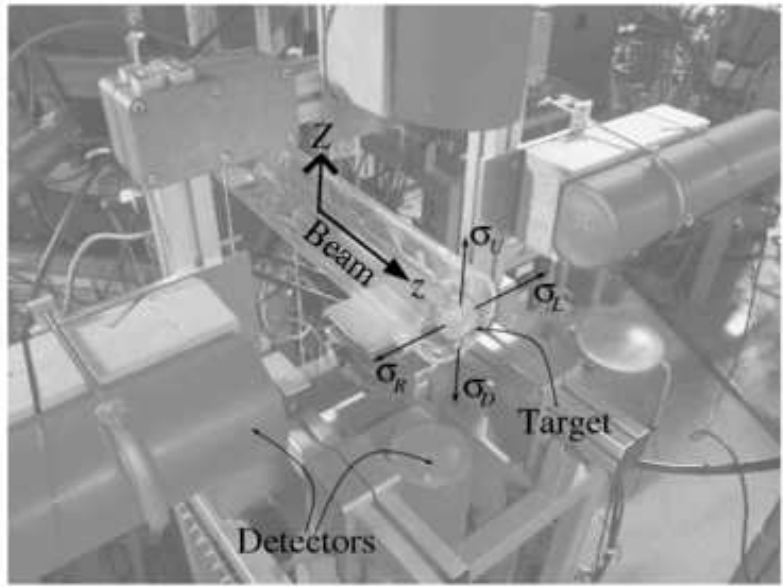}}
\caption{\small{Photograph of the LDP showing beam tube, target
holder and four neutron detectors. Also shown are the directions of
the beam, the quantization axis, and left, right, up, and down
scattering.}} \label{fig:setup}
\end{figure}

The LDP consists of a target and four detectors as shown in
Fig.\,\ref{fig:setup}. A polarized deuteron beam from POLIS impinges
on a target containing deuteron in the form of C$_2$D$_4$ to reacts
as D($\vec d, n$)$^{3}$He sending neutrons in space.

\subsubsection{Target} A deuterated-polyethylene (C$_2$D$_4$) film
was used as a deuterium target. It has the same chemical properties
as polyethylene (i.e. the hydrocarbon C$_2$H$_4$), except that it
contains deuterium nuclei (deuterons) instead of hydrogen nuclei
(protons). Although carbon is present in the target material, it
does not produce neutrons in a nuclear reaction because the Q-value
of the $^{12}$C($d,n$)$^{13}$N reaction ($Q= -281$ keV) is higher
than our beam energy. Therefore neutrons detected originate
exclusively from the D($d, n$)$^{3}$He reaction. Alternative targets
to C$_2$D$_4$ exist, such as, deuterated-titanium, or deuterium gas
targets. These were not used because C$_2$D$_4$ targets were readily
available at the KVI and are more convenient to produce. Target
thickness was in the order of a few hundred $\mu$g/cm$^2$, which
corresponds to a target depth of
a few $\mu$m.\\

At early stages of the experiment, targets consisted of a C$_2$D$_4$
thin film held by a rectangular frame. Eventually targets evolved
into a C$_2$D$_4$ thin film on a round metal backing. There were two
problems with the initial target design. First, the target would
melt under beam heating. The reason was that Polyethylene, being a
plastic, conducts poorly the energy deposited by the beam. The
solution was to couple the thin-film to a metallic backing acting as
a heat sink. Secondly, the sharp edges of the rectangular frame help
produce unwanted electrical discharge when the target was at High
Voltage. A round metal holder increased the breakdown voltage. The
breakdown voltage\footnote{A high voltage electrode in pressures of
10$^{-2}$ to 10$^{-3}$ mbar generates all kinds of plasma effects
that are beautiful to watch, such as, striations and
micro-discharges.} was found to be proportional to the pressure in
the evacuated beam line, the deuteron beam current, and surface
conditions on the target.

\subsubsection{Detectors} The detectors we used were `liquid
organic scintillators' of type NE213. These detectors are frequently
used for the detection of neutrons. The signal produced by these
detectors depends on the
type of particle entering them.\\

\subsubsection{Pulse Shape Discrimination}

\begin{figure}[]
\hskip 0.3cm \scalebox{0.9}[0.9]{\includegraphics*[viewport=0 0 375
200]{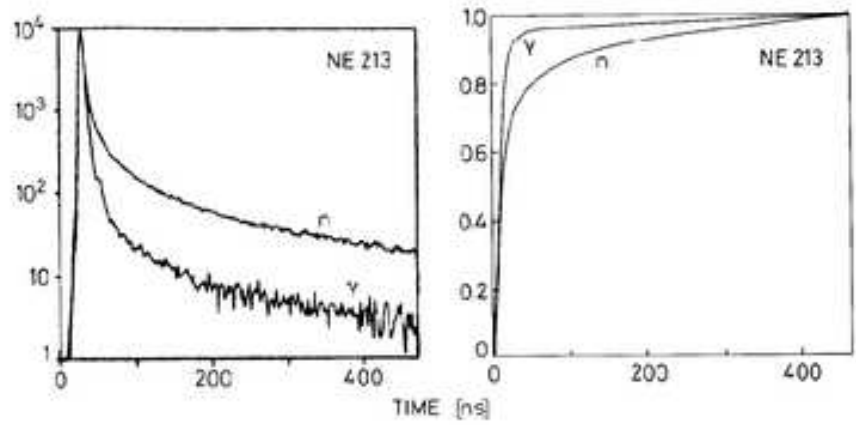}} \caption{\small{Pulse shape differences for
neutron and $\gamma$-rays. Detector signals are shown on the right
panel and the integral of the signals on the left panel.}}
\label{fig:sigshape}
\end{figure}

As determined earlier (Sec.\,\ref{sec:stoppingpower} and
\ref{sec:d+dreaction}), only neutrons and $\gamma$-rays will reach
the detectors. One wants to distinguish between detected neutrons
and $\gamma$-rays because one needs to measure the asymmetry
originating from a single reaction, and not two reactions. In our
case neutrons should be counted, while $\gamma$-rays should be
rejected. A method called Pulse Shape Discrimination \cite{leo,
heltsley} allows different particles to be distinguished based on
the signal they produce in detectors. Different particles have
different energy loss mechanisms inside matter, and so produce
sightly different signal shapes. Neutrons will deposit their energy
more slowly than gamma-rays, therefore the signal they create decays
more slowly. The signals are illustrated in
Fig.\,\ref{fig:sigshape}. \\

\begin{figure}
\hskip2cm\scalebox{0.35}[0.35]{\includegraphics[]{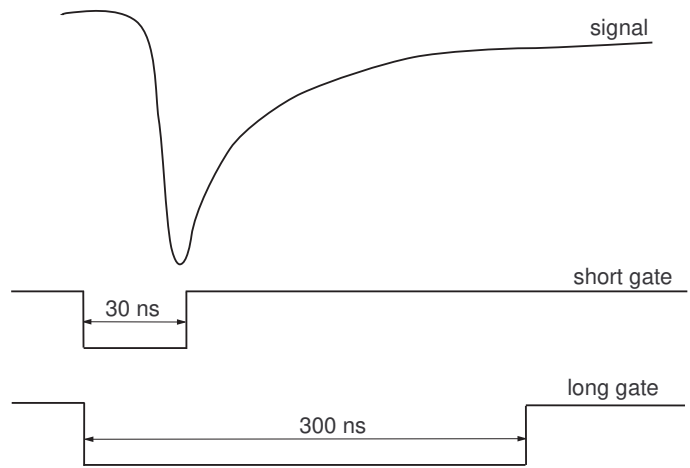}}
\caption{\small{Detector signal and integration gates as used in our
experiment.}} \label{fig:psdgates}
\end{figure}

\begin{figure}[!h]
\hskip2.5cm \scalebox{0.5}[0.45]{\includegraphics[]{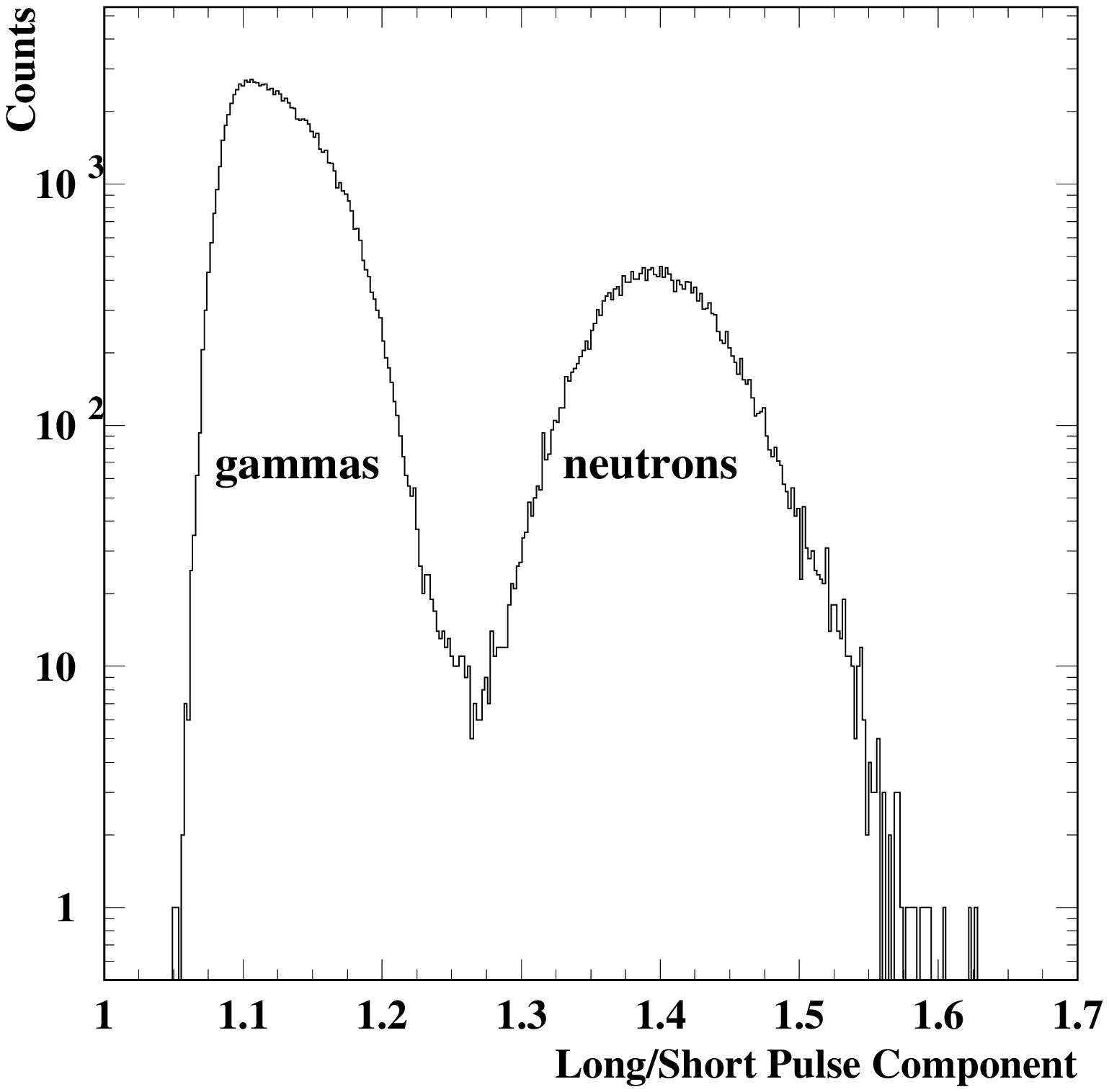}}
\caption{\small{Pulse shape spectrum for one detector. Neutrons are
well separated from a large background of $\gamma$-rays. The number
of neutrons detected is the number of events under the neutron
peak.}} \label{fig:1psd}
\end{figure}

By monitoring the signal shape which the detected particles produce,
one can identify particles. We monitored the signal shape by
integrating two copies of the signal with two different time
intervals, and taking their ratio. The signal and time intervals are
illustrated in Fig.\,\ref{fig:psdgates}.
The ratio of these two integrated quantities is proportional to the
decay time of the signal, and is the basis for particle
identification. By plotting the occurrence of signals as a function
of this ratio we get a pulse shape spectrum as shown in
Fig.\,\ref{fig:1psd}.\\

\begin{figure}[!h]
\hskip 2cm \scalebox{0.5}[0.5]{\includegraphics[]{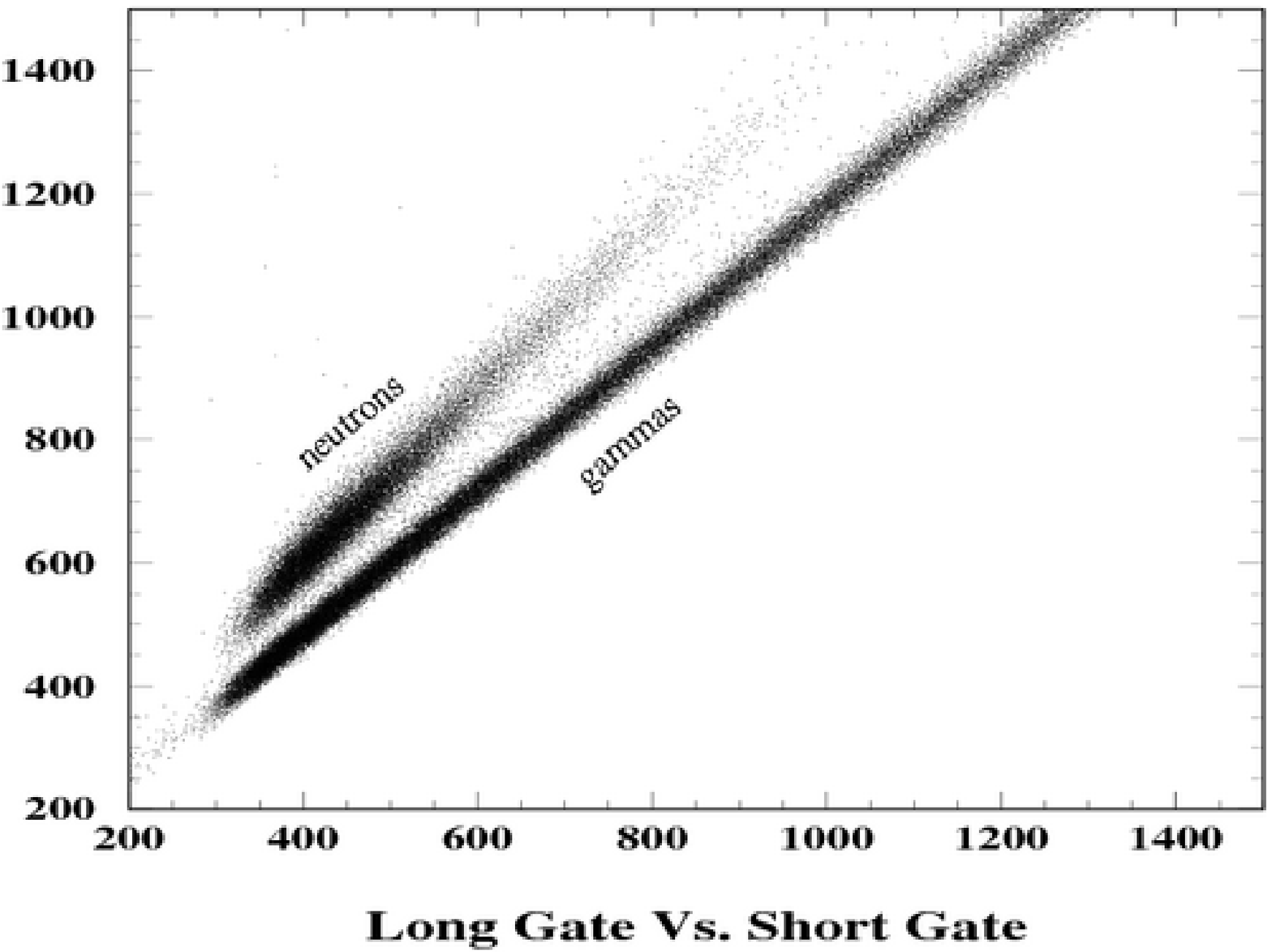}}
\caption{\small{2-D pulse shape spectrum.  Detected neutrons can
clearly be separated from detected $\gamma$-rays.}}
\label{fig:2dpsd}
\end{figure}

We can also plot the integral of the signal with the long gate
versus that with the short gate to get a 2-D scatter plot of pulse
shapes, as shown in Fig.\,\ref{fig:2dpsd}.\\

\subsubsection{Post Acceleration} Placing the target at a potential
is an experimental trick to reduce the measuring time of our
experiment. The cross section of the D$(d,n)^3$He reaction was found
to depend on the reaction energy (Fig.\,\ref{fig:crsd}). By
increasing the energy of the beam, the cross section, therefore
reaction rate, increases. High counting rates have the obvious
advantage of lower statistical uncertainty for a given measuring
time. The maximum beam energy that POLIS can produce is 30 keV.
Placing the target at negative potentials will accelerate the
positively charged deuteron beam. If the beam is initially at 30 keV
and the target at $-50$ kV, then the beam energy as it reaches the
target is 80 keV. The reaction rate at 80 keV is about an order of
magnitude
higher than at 30 keV!\\

Post acceleration equipment is not shown in the photograph of the
LDP (Fig.\,\ref{fig:setup}). It consists of nothing more than a high
voltage cable connecting the target to a high voltage power supply
surrounded by safety features (insulation and grounding).

\subsubsection{Data Acquisition} Data acquisition consists of
electronic modules, a CAMAC crate, and a PC. Electronic modules can
manipulate and perform basic operations on electronic signals. The
detector signal was first split into multiple copies by a
`fan-in/fan-out' unit. Two copies of the signal were each put into a
`constant fraction discriminator' (CFD) unit to generate logical
gates. One long and one short gate acts as integration windows for
the detector signal as shown in Fig.\,\ref{fig:psdgates}. Another
copy of the signal was cable delayed before being put to a `charge
to digital converter' (QDC) units along with the short and long
gate. The QDC integrates the voltage of a signal and converts it to
into a binary string which can be read by a PC. The program DAX,
based on the CERN Program Library package \cite{cern}, was used to
write raw data into event files (*.ntuple). PAW \cite{paw} was used
to perform a simple analysis of the data such as plotting and
counting.

\newpage
\section{Data Analysis}

\subsection{Low-Energy Deuteron Polarimeter}
\subsubsection{Analyzing Powers}
Analyzing powers are a manifestation of the dependence of the
reaction cross section on spin. The analyzing powers relevant to
this experiment are almost entirely available in the literature.
Becker \cite {becker} has measured all analyzing powers ($A_y$,
$A_{xz}$, $A_{zz}$, $A_{xx}-A_{yy}$) for the D$(d, n)^3$He and
D$(d,p)^3$H reactions at the reaction energy of $E=28$ keV. Fletcher
\cite{fletcher, fletcherphd} has measured two tensor analyzing
powers ($A_{zz}$, $A_{xx}-A_{yy}$) of both reactions at the beam
energies of $E_d=$ 25, 40, 60, and 80 keV. Tagashi \cite{tagashi}
has measured all analyzing powers of the D$(d,p)^3$H reaction at
$E_d=$ 30, 50, 70, and 90 keV. Our polarimeter is limited to the
energy range of $E_d=$ 25 to 80 keV, because that is the energy
range at which analyzing powers are presently known, also because
those are the beam energies
available to us.\\

$A_{xz}$ and the energy dependence of $A_y$ are not published. $A_y$
is claimed \cite{becker} to be energy independent in our energy
range. Furthermore, $A_y$, for the D$(d,p)^3$H reaction, has
negligible energy variation as can be seen from \cite{tagashi}. An
unknown $A_{xz}$ is not a problem for us since we do not use it in
our analysis. When $\beta=90^{\circ}$ (as in our setting of the
beam) only $\epsilon_1$, $\epsilon_3$, and
$\epsilon_5$ enter the analysis (Eq.\,\ref{eq:spin1asym}).\\

\begin{figure}[!h]
\hskip1cm \scalebox{0.9}[0.9]{\includegraphics[]{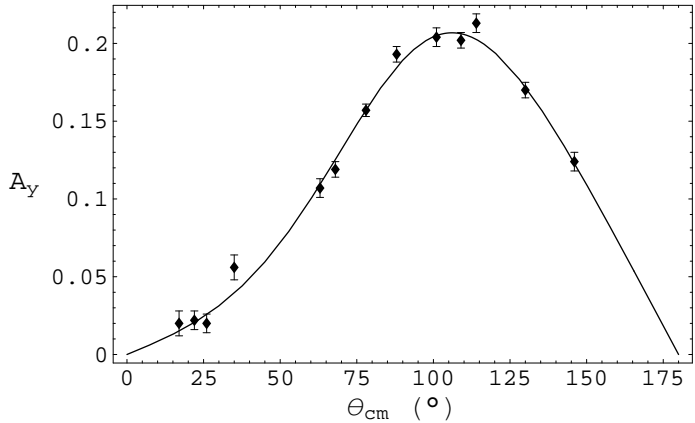}}
\caption{\small{Data and fit of $A_y$ at 28 keV from Becker
\cite{becker}.}} \label{fig:ay}
\end{figure}

\begin{figure}[!h]
\hskip1cm \scalebox{1}[1]{\includegraphics[]{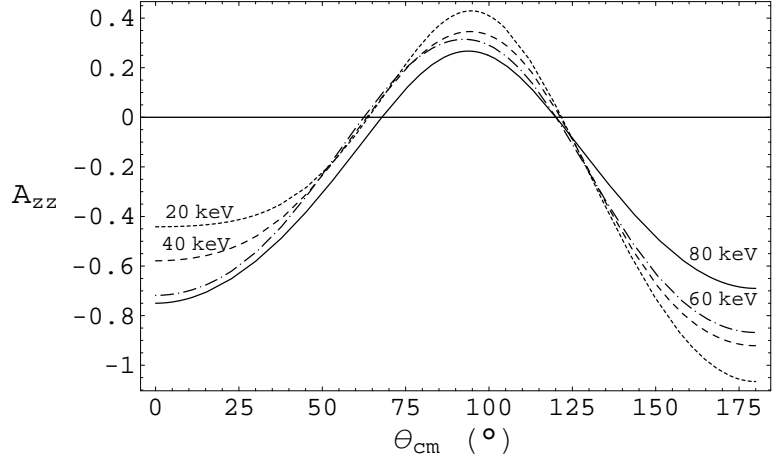}}
\caption{\small{Fit of $A_{zz}$ at various beam energies from
Fletcher \cite{fletcherphd}.}} \label{fig:azze}
\end{figure}

 Analyzing powers are usually reported in the literature
as data points with fitted curve (Fig.\,\ref{fig:ay}) as in
\cite{becker, fletcherphd}, or just fitted curves
(Fig.\,\ref{fig:azze}) as in \cite{fletcher, tagashi}. The fitting
functions are Legendre polynomials.

\subsubsection{Effective Analyzing Powers}
In our analysis one cannot directly use the analyzing powers as
found in the literature. The asymmetries measured by the detectors
are the convolution of both non-zero analyzing powers and
experimental effects, such as:

\begin{enumerate}
\begin{small}
    \item Finite acceptance of the detectors. Detectors cover
    a non-zero solid angle. Therefore, the analyzing power must be
     averaged over this solid angle. We chose to average $A(\theta)$
     over $\Delta\theta$, and disregard $\Delta\phi$ effects because
      they are small at $\theta_{lab}=90^{\circ}$.
    \item Change in analyzing power due to the energy loss of the beam
    through the target, $A(E)$. Fig.\,\ref{fig:azze} shows that $A_{zz}$ is energy dependent.
    \item Change in cross section due to energy loss of the beam
    though the target, $\sigma(E)$.
    \item Energy profile of the beam through the target; $E(x)$.
\end{small}
\end{enumerate}

These experimental effects give rise to an effective analyzing power
given by

\begin{equation}\label{eq:effectv}
A^{eff}= \frac{\int_{\Delta\theta}
 \int_{\Delta x} A\big(E(x),\theta\big) \,\, \sigma \big(
E(x)\big)\, d\theta  \, dx}{\Delta\theta\ \int_{\Delta x} \sigma
\big( E(x)\big) \, dx}.
\end{equation}

The opening angle of each of our detectors is
$\Delta\theta=26^{\circ}$. $\Delta x$ is the range of the beam.
$E(x)$ and $\Delta x$ can be calculated from the stopping power of a
deuteron beam on a C$_2$D$_4$ target (Eq.\,(\ref{eq:empdedx}),
$E(0)=E_d$ and $E(range)=0$). $A(E,\theta)$ represents the energy
and polar angle dependence of any of the four analyzing powers
($A_y$, $A_{xz}$, $A_{zz}$, $A_{xx}-A_{yy}$). $A(\theta$) is given
in the literature. $A(E)$ was estimated by fitting a quadratic
polynomial
through the four energies at which analyzing powers are known.\\

The effective analyzing powers at two different
beam energies are given in Table\,\ref{tab:effa}.\\

\begin{table}
\begin{center}
\begin{small}
\begin{tabular} {|c||c|c|}
\hline
 & $E_d=$ 50 keV &$E_d=$ 80 keV\\
\hline\hline
 $A_y^{eff}$&$0.186 \pm 0.01$&$0.186 \pm 0.01$\\
 \hline
$A_{zz}^{eff}$&$0.31 \pm 0.03$& $0.25 \pm 0.03$\\
 \hline
 $(A_{xx}-A_{yy})^{eff}$&$-0.98 \pm 0.1$&$-0.83 \pm 0.1$\\
\hline
\end{tabular}
\end{small}
\end{center}
\caption{\small{Effective analyzing power of the LDP for 50 and 80
keV beams.}} \label{tab:effa}
\end{table}

\subsubsection{Instrumental Asymmetries}

\begin{figure}[h]
\hskip1cm\scalebox{0.6}[0.6]{\includegraphics[]{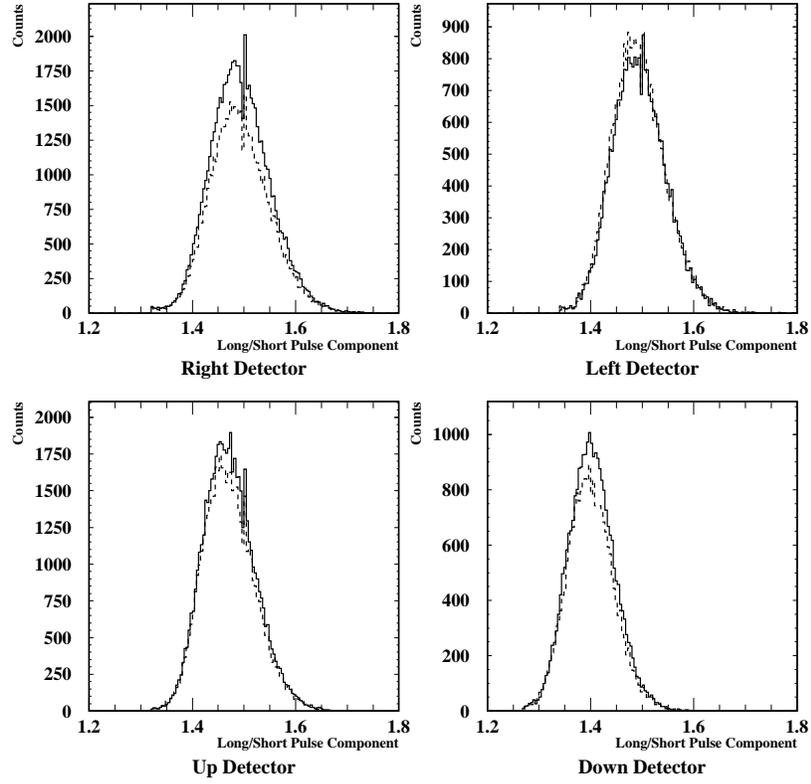}}
\caption{\small{Neutron peak in each of the LDP's detectors. The
dashed line stems from an unpolarized beam, whereas the solid line
stems from a positive vector polarized beam.}} \label{fig:4psd}
\end{figure}

Fig.\,\ref{fig:4psd} shows the neutron peaks in each detector coming
from the reaction with a polarized and unpolarized beam.

One can observe that the neutron peaks with an unpolarized beam are
not of the same height, indicating asymmetries! For example, the
peak of the Right and Up detector approach 2000 neutrons per bin,
while the Left and Down peaks approach 1000 neutrons. This is due to
instrumental asymmetries. One should distinguish between reaction
asymmetries and instrumental asymmetries. Reaction asymmetries are
fundamental and due to a non-zero analyzing power of the nuclear
reaction. Instrumental asymmetries are due to mismatches in detector
response, such as differences between detectors in, gain, solid
angle, detection efficiency, discriminator threshold level, or
signal shape. The fact that instrumental asymmetries are equal to a
factor of 2 here, indicates that the detector responses were not
matched properly. This is not a problem because one uses unpolarized
beams to quantify instrumental asymmetries. Unpolarized beams
theoretically do not create reaction asymmetries. Therefore, any
asymmetry measured with unpolarized beams is taken to be
instrumental asymmetries. Normalizing the detector counts obtained
with polarized beams, by counts obtained from unpolarized beams,
leaves only the reaction asymmetries needed to determine beam
polarization. Therefore, for the right detector, the number of
neutrons creating reaction asymmetries is,

\[ R=\frac{R_{polarized}}{R_{unpolarized}}.\]
And similarly for the other three detectors $L$, $U$, and $D$.\\

Until this point, we have covered the effective analyzing powers and
the normalized asymmetries. Using Eq.\,(\ref{eq:spin1asym}), one
solves a system of equations to get the unknowns ($P_Z$, $P_{ZZ}$).

\subsubsection{Sample Calculation}

As an example of the procedure to calculate the polarization from
neutron counts, we analyze the data presented in
Fig.\,\ref{fig:4psd}.

\begin{eqnarray*}
  R &=& {62566}/{51124}=1.22380, \\
  L &=& {25815}/{26985}=0.95664, \\
  U &=& {56244}/{51337}=1.0955,\\
  D &=& {26600}/{23455}=1.1340.
\end{eqnarray*}

With the asymmetries,

\begin{eqnarray*}
  \epsilon_1 &=& \frac{L-R}{L+R} = 0.1225 =\frac{0.279\,P_Z\sin\beta}
  {1+\frac{1}{2}P_{ZZ}[-\sin^2\beta\,0.647+\cos^2\beta\,0.312]} \\
  \epsilon_2 &=&  \frac{U-D}{U+D} = -0.0172 = \frac{0.0109\,P_{ZZ}\sin\beta
\,\cos\beta}{1+\frac{1}{2}P_{ZZ}[\sin^2\beta\,0.958+\cos^2\beta\,0.312]} \\
  \epsilon_3 &=& \frac{2(L-R)}{L+R+U+D} = 0.1211= \frac{0.279\,P_Z\sin\beta
  }{1+0.078\,P_{ZZ}[3(\cos^2\beta-1)\,0.312]} \\
  \epsilon_4 &=& \frac{2(U-D)}{L+R+U+D} = -0.0174 =
\frac{0.0109\,P_{ZZ}\sin\beta \,\cos\beta}{1+0.078\,P_{ZZ}[3(\cos^2\beta-1)\,0.312]} \\
  \epsilon_5 &=& \frac{(L+R)-(U+D)}{L+R+U+D}= -0.0111 =
\frac{0.245\,P_{ZZ}\sin^2\beta \,\,
}{1+0.078\,P_{ZZ}[3(\cos^2\beta-1)\,0.312]}
\end{eqnarray*}

\begin{figure}[!h]
\hskip0cm \scalebox{0.9}[0.9]{\includegraphics[]{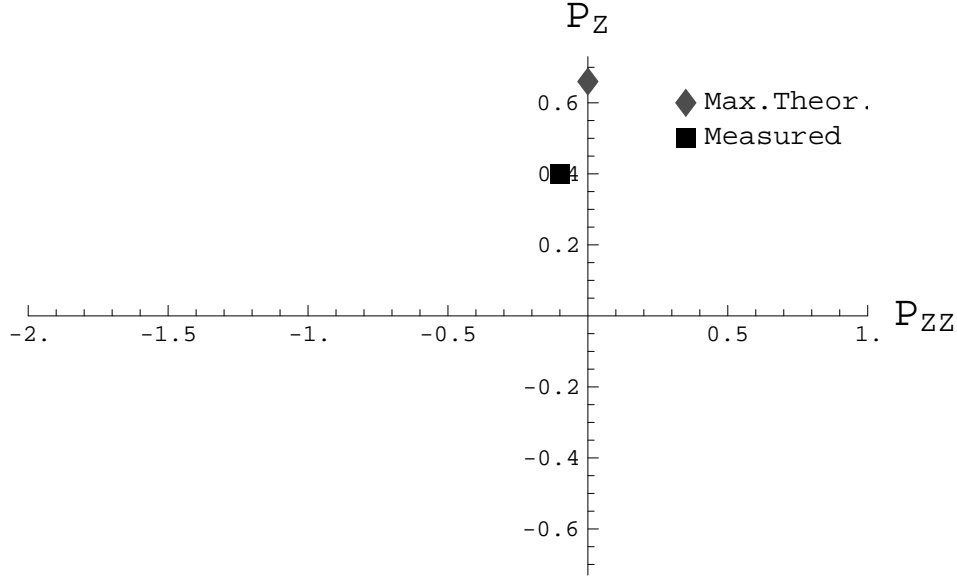}}
\caption{\small{Measured polarization and maximum theoretical
polarization for a positive vector polarized beam. The vertical axis
is the vector component of polarization, while the horizontal axis
is the tensor component. In this plot vector polarized beams lie on,
or close to, the vertical axis, while tensor polarized beams lie on,
or close to, the horizontal axis.}} \label{fig:samplefig}
\end{figure}

Fixing $\beta=90^\circ$ and solving for $P_{ZZ}$ in $\epsilon_5$
yields $P_{ZZ}= -0.05$. Solving for $P_Z$ in $\epsilon_3$ or
$\epsilon_1$ using $P_{ZZ}$ yields $P_Z=0.4$. Applying propagation
of errors to these equations with the statistical uncertainty
originating from counting uncertainty ($\Delta N=\sqrt{N})$, and the
systematic uncertainty originating from the uncertainty in analyzing
powers given in Table \ref{tab:effa}, yields, $\Delta
P_{ZZ}^{stat}=0.001$, $\Delta P_Z^{stat}=0.003$, $\Delta
P_{ZZ}^{sys}=0.005$, and $\Delta P_{Z}^{sys}=0.02$. This measurement
of polarization is plotted in Fig.\,\ref{fig:samplefig}.

\subsubsection{Low Energy Deuteron Polarimeter}

\begin{figure}[!h]
\hskip0cm \scalebox{0.9}[0.9]{\includegraphics[]{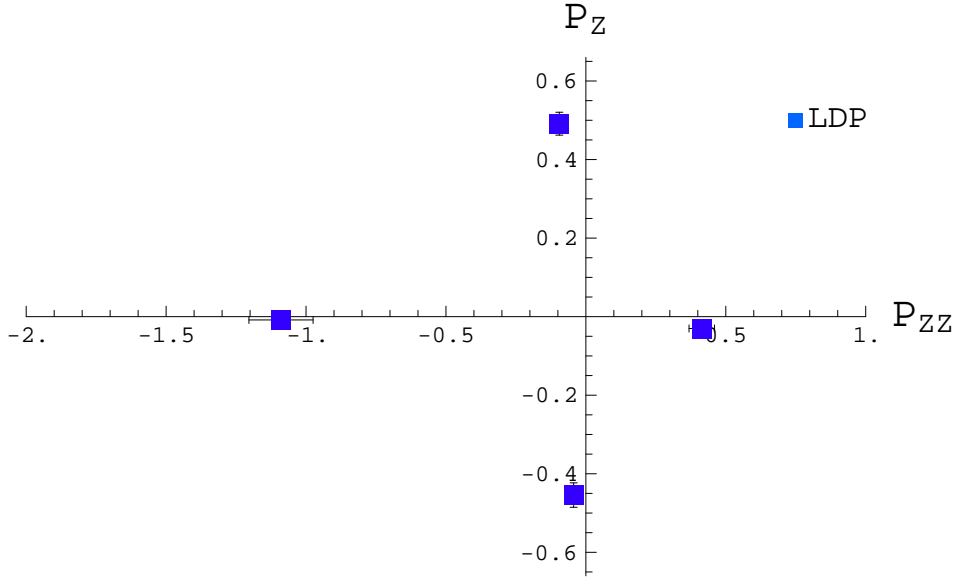}}
\caption{\small{LDP measurements of polarization of four polarized
beams. Systematic error bars are just visible, while statistical
error bars are smaller than the symbol size.}} \label{fig:ldpfig}
\end{figure}

The Low Energy Deuteron Polarimeter (LDP) data are tabulated in
Table\,\ref{tab:ldptab} and plotted as squares in
Fig.\,\ref{fig:ldpfig}.

\begin{table}[!h]
\hskip -0.8 cm
\begin{scriptsize}
\begin{tabular}{|c||c|c|c|c||c|c|}
 \hline
 POLIS &\multicolumn{4}{c||}{Detector}&\multicolumn{2}{c|}{Measured}\\
 State &\multicolumn{4}{c||}{Count}&\multicolumn{2}{c|}{Polarization}\\
\cline{2-7}&R&L&U&D&$P_Z \stackrel{stat}{\pm} \,
\stackrel{sys}{\pm}$&
$P_{ZZ} \stackrel{stat}{\pm} \, \stackrel{sys}{\pm}$\\
\hline\hline
 WF & 30703 &43889 &22943& 51843 &$0.49 \pm
 0.0008 \pm 0.03 $ &$-0.095\pm 0.003 \pm 0.005$\\
 ST. I + ST. II&25815&62566&26600&56244&$-0.45 \pm
 0.001 \pm 0.03 $ &$-0.043 \pm 0.003 \pm 0.0006$\\

 MF + ST. I&29379& 53164 &20200& 41951&$-0.030 \pm
 0.001 \pm 0.005 $ &$0.414 \pm 0.003 \pm 0.05$\\

 MF + ST. II&18011& 33064 &24853&54641&$-0.008 \pm
 0.0009 \pm 0.002 $ &$-1.1 \pm 0.004 \pm 0.1$\\
 Unpolarized &54312&100427&444617&101696&0&0\\
\hline
\end{tabular}
\end{scriptsize}
\caption{\small{LDP counts and polarizations.}}\label{tab:ldptab}
\end{table}

\subsection{In-Beam Polarimeter}
The In-Beam Polarimeter (IBP) measures similarly to the LDP. Namely,
by measuring asymmetries and exploiting known analyzing powers. The
reaction used to measure asymmetries was H($d,d$)p at $E_d=$ 80 MeV.
The analyzing powers for this reaction, at this energy, are not
reported in the literature, so we used calculated \cite{arnoldas}
analyzing powers shown in Table \,\ref{tab:aibp}. The uncertainties
in analyzing power ($\approx 3\%$) were taken as the variation in
analyzing power between
various potentials.\\

\begin{figure}[!h]
\hskip0cm \scalebox{0.9}[0.9]{\includegraphics[]{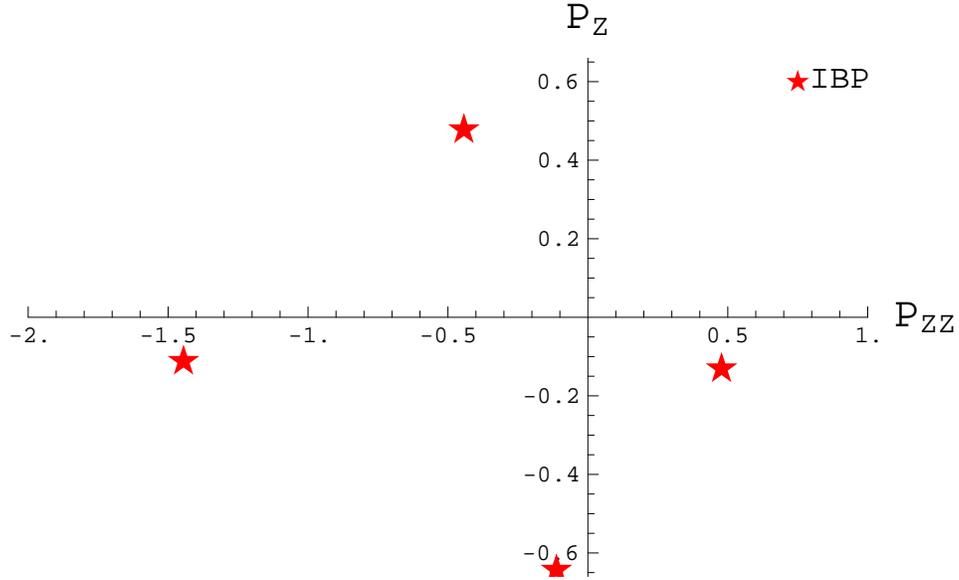}}
\caption{\small{IBP measurement of polarization of four polarized
beams. Error bars are not visible.}} \label{fig:ibpfig}
\end{figure}

The IBP data are tabulated in Table\,\ref{tab:ibptab} and plotted as
stars in Fig.\,\ref{fig:ibpfig}.

\begin{table}[!h]
\begin{center}
\begin{small}
\begin{tabular} {|c|c|}
\hline
 & $E_d=$ 80 MeV \\
  & $\theta_{cm}=100^{\circ}$\\
\hline
 $A_y$&$-0.291 \pm 0.005$\\
\hline
$A_{zz}$&$-0.74 \pm 0.02$\\
\hline
 $A_{xx}-A_{yy}$&$-0.107\pm 0.003$\\
\hline
\end{tabular}
\end{small}
\end{center}
\caption{\small{IBP analyzing powers.}} \label{tab:aibp}
\end{table}

\begin{table}[!h]
\hskip -0.8 cm
\begin{scriptsize}
\begin{tabular}{|c||c|c|c|c||c|c|}
 \hline
 POLIS &\multicolumn{4}{c||}{Detector}&\multicolumn{2}{c|}{Measured}\\
 State &\multicolumn{4}{c||}{Count}&\multicolumn{2}{c|}{Polarization}\\
\cline{2-7}&R&L&U&D&$P_Z \stackrel{stat}{\pm} \,
\stackrel{sys}{\pm}$&
$P_{ZZ} \stackrel{stat}{\pm} \, \stackrel{sys}{\pm}$\\
\hline\hline
 WF & 29589 &17220 &25004& 24368 &$0.478    \pm 0.0008 \pm 0.01
   $&$ -0.443  \pm 0.006\pm 0.001   $\\
ST. I + ST. II&13180&20140&18465&18085&$-0.643   \pm 0.001 \pm 0.008
   $&$ -0.112   \pm   0.005  \pm 0.004   $\\
MF + ST. I&19407& 17626 &18181&17932&$-0.131   \pm 0.001 \pm 0.003
   $&$ 0.478  \pm  0.007 \pm 0.004   $\\
MF + ST. II&10834& 10257 &13681&13391&$-0.112  \pm 0.001
\pm 0.002 $&$ -1.44  \pm 0.009  \pm 0.01   $\\
Unpolarized &71690&73594&75327&75093&0&0\\
\hline
\end{tabular}
\end{scriptsize}
\caption{\small{IBP counts and polarizations.}}\label{tab:ibptab}
\end{table}

\newpage
\subsection{Lamb-Shift Polarimeter}

The Lamb-Shift Polarimeter (LSP) uses a different technique, than
the LDP and IBP, to measure polarization. It measures directly the
spin substate distribution of a beam, from which the polarization
can easily be obtained.\\

\begin{figure}[!h]
\hskip0.5cm
\includegraphics[]{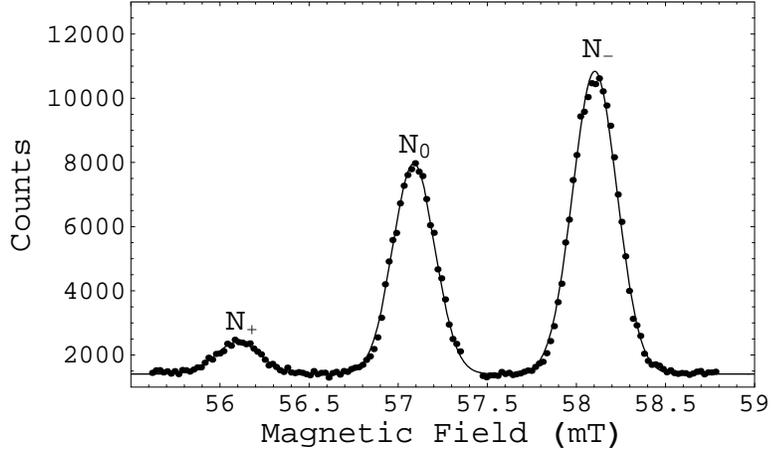}
\caption{\small{Spin state distribution of a negative vector
polarized beam measured by the LSP.}} \label{fig:lsp}
\end{figure}

Let $N_+$ be the population of spins in the spin substate $m_I=+1$.
$N_0$ the population of spins in the $m_I=0$ state, and $N_-$ for
$m_I=-1$. The polarization in terms of populations is,

\begin{eqnarray*}
  P_Z &=& \frac{N_+ - N_-}{N_++ N_0+N_-} \\
  P_{ZZ} &=& 1-3\frac{N_0}{N_++ N_0+N_-}.
\end{eqnarray*}

Populations are calculated by fitting three gaussian peaks and a
baseline to the LSP's spin substate distribution, as is illustrated
in Fig.\,\ref{fig:lsp}. The area under the peaks are equal to the
populations $N_+$, $N_0$, and $N_-$. The uncertainty in fitting
parameters (obtained from the fitting program) were propagated into
uncertainties in polarization. The data shown in Fig.\,\ref{fig:lsp}
yield $P_{ZZ}=-0.121  \pm 0.004$ and $P_Z=-0.51 \pm 0.01.$ These
uncertainties are purely
statistical.\\

\begin{figure}[!h]
\hskip0cm \scalebox{0.9}[0.9]{\includegraphics[]{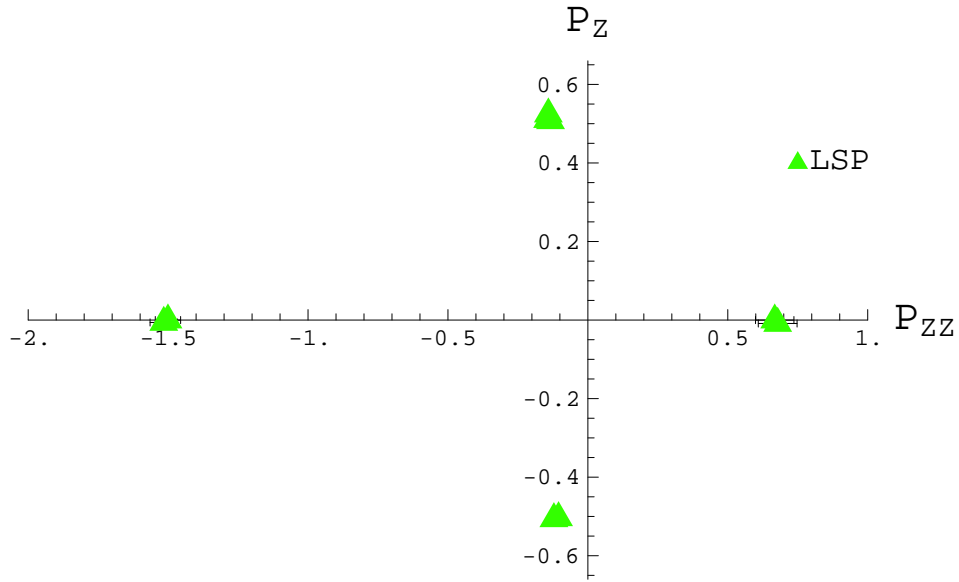}}
\caption{\small{LSP measurement of polarization of four polarized
beams.}} \label{fig:lspfig}
\end{figure}

The LSP data are tabulated in Table\,\ref{tab:lsptab} and plotted as
triangles in Fig.\,\ref{fig:lspfig}.

\begin{table}[!h]
\begin{small}
\begin{center}
\begin{tabular}{|c||c|c|}
 \hline
 POLIS &$P_Z$&$P_{ZZ}$\\
 State & & \\
\hline\hline
WF  &$ -0.51 \pm 0.01 $&$-0.121\pm0.004$\\
WF  &$ -0.50 \pm0.01 $&$-0.105\pm0.003$\\
ST. I + ST. II&$0.52\pm 0.01 $&$-0.141\pm 0.005$ \\
ST. I + ST. II &$ 0.510 \pm 0.01 $&$-0.144\pm0.004$\\
ST. I + ST. II & $0.507 \pm 0.003 $&$-0.133\pm0.004$\\
MF + ST. I &$ -0.009 \pm 0.009 $&$0.68\pm0.07$\\
MF + ST. I&$0\pm0.07$&$0.67\pm0.07$ \\
MF + ST. II & $ 0 \pm 0.01 $&$-1.50\pm0.05$\\
MF + ST. II & $ 0 \pm 0.02 $&$-1.52\pm0.05$\\
\hline
\end{tabular}
\end{center}
\end{small}
\caption{\small{LSP data, taken within a few minutes of each
other.}}\label{tab:lsptab}
\end{table}

\newpage
$ $ 
\newpage

\section{Results}
\begin{figure}[!h]
\hskip0cm \scalebox{0.9}[0.9]{\includegraphics[]{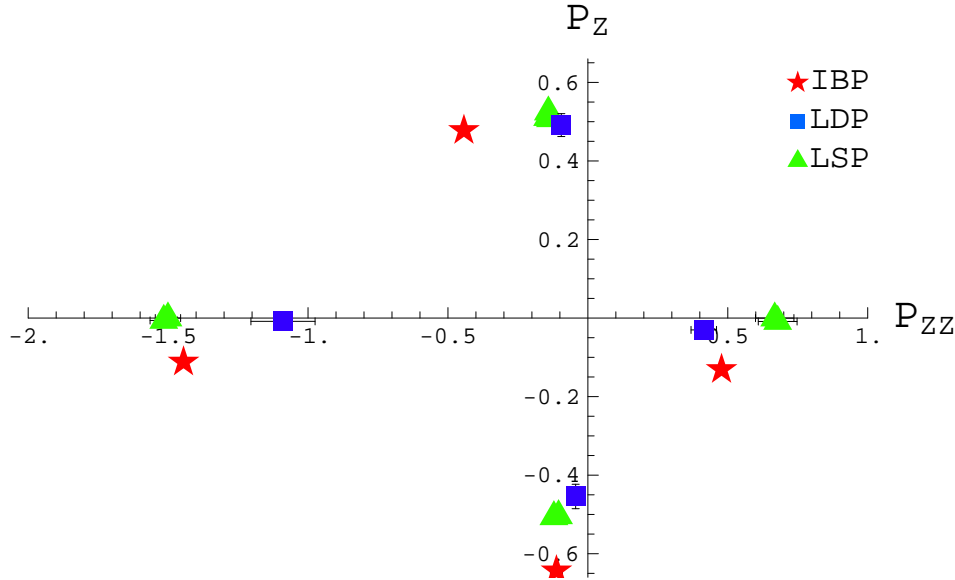}}
\caption{\small{Polarization measurements of all three
polarimeters.}} \label{fig:polresult}
\end{figure}

The measurements taken by the three polarimeters in
Fig.\,\ref{fig:ldpfig}, Fig.\,\ref{fig:ibpfig}, and
Fig.\,\ref{fig:lspfig} are superimposed in
Fig.\,\ref{fig:polresult}.

These data were collected on the night of 15-16 October 2004. LSP
measurements were taken between 10 and 11 pm, IBP measurements were
taken between 3 and 5 am, and LDP measurements were taken between 6
and 11 am. The LSP and LDP disagree on the tensor component of
polarization, while the LSP and IBP disagree on the vector and
tensor components of polarization. It is difficult to draw
definitive conclusions from this plot for three reasons:

\begin{enumerate}
\item The LDP does not fully agree either with the IBP or LSP, not allowing one to
prove whether the polarization changes between LSP and IBP,
\emph{or} if the LSP or IBP have unknown systematic errors.
\item We do not have a big data set with multiple measurements at this moment. \item We cannot exclude
changes in beam polarization while the polarimeters were being
switched. Had we monitored the polarization with one polarimeter
(LSP) before and after the measurements of other polarimeters, we
would know whether the beam polarization was constant over the
measurement period.
\end{enumerate}

Repeating the experiment would have been the natural way to proceed,
this would have addressed points 2 and 3. Yet because of time and
facility constraints, we have to evaluate the present data.\\

The IBP data are not reliable for two reasons. Firstly, we do not
know how successful the theory, which calculates IBP's analyzing
powers, is. Secondly, the asymmetries we measured depended on how we
analyzed the data (where we placed cuts). Both of these issues can
be addressed with more time. Until then we cannot be confident about
the present IBP polarization measurements and uncertainty in
polarizations. Therefore we cannot presently answer the question
whether there are polarization
changes in the accelerator.\\

One can observe that the polarization measurements of the LSP and
LDP agree quite well for the $P_Z$ component of polarization but not
for $P_{ZZ}$. One can also see that the LDP measures always less
$P_{ZZ}$ than the LSP (all LDP points are closer to the vertical
axis in Fig.\,\ref{fig:polresult}).\\

We can rule out large decreases of $P_{ZZ}$ in time between the LSP
and LDP measurements. If the polarization were to change, it must do
so in particular ways. If the dissociator of POLIS were to operate
anomalously (a decrease in efficiency), then the polarization of
both vector and tensor polarized beams would decrease. This is not
observed since only the tensor polarized beams have lower
polarization. If the medium field transitions units of POLIS (the
transition field allowing tensor polarized beams) would operate
anomalously, then the polarization would vary diagonally on a
$P_{ZZ}$ vs. $P_Z$ plot. This is not seen since, $P_Z$ of tensor
polarized beams is the same for both the LSP and LDP, ruling out
variations in the medium field transition unit. We are left with the
conclusion that the LDP
underestimates the tensor component of beam polarization.\\

\begin{figure}[!h]
\hskip0cm \scalebox{0.9}[0.9]{\includegraphics[]{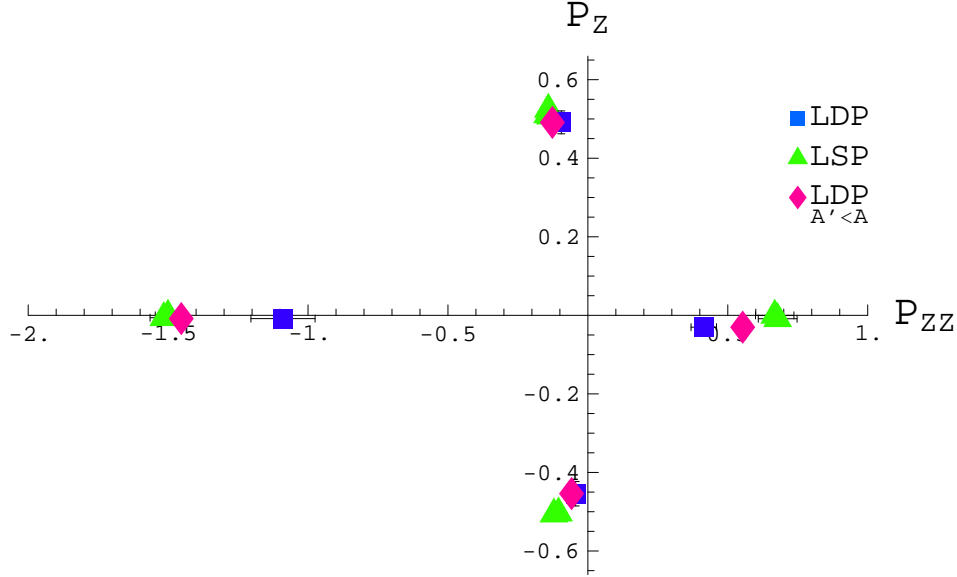}}
\caption{\small{LDP with reduced tensor analyzing powers (diamond
symbol) is in better agreement with the LSP.}} \label{fig:ldpprime}
\end{figure}

The agreement between LSP and LDP can be improved if one assumes
lower tensor analyzing powers. By reducing the tensor analyzing
powers ($A_{xx}-A_{yy}$ and $A_{zz}$) of the LDP by 25\%, $P_{ZZ}$
increases as shown in Fig.\,\ref{fig:ldpprime}. This suggest that
the effective tensor analyzing powers ($(A_{xx}-A_{yy})^{eff}$ and
$A_{zz}^{eff}$) we used are wrong. Either because our model
(Eq.\,\ref{eq:effectv}) for effective tensor analyzing power is
incomplete\footnote{We did not include the effect of beam
straggling, nor the angular efficiency profile of the detectors, nor
the effect of material coating the front of the target.}, or because
the analyzing powers we used in the model were wrong. We used the
tensor analyzing powers published by Fletcher \cite{fletcher} for
two reasons. Firstly because they were closer to the theoretical
value predicted. Secondly because they were published for a number
of energies, allowing us to estimate the energy dependence of the
analyzing powers, and use beams of any energy between 25 and 80 keV.
The tensor analyzing powers published by Becker \cite{becker} were
not used because they were inconsistent with the theoretical
predictions, and also because they were published at a single
energy. The difference in tensor analyzing power between these two
publications is considerable. For example, at $E_d\approx30$ keV and
$\theta_{cm}=90^{\circ}$, Fletcher \cite{fletcher} has
$A_{xx}-A_{yy}$ and $A_{zz}$ approximately 35\% and 50\% greater
than Becker \cite{becker}. Our effective analyzing powers were
derived from Fletcher's analyzing powers, yet our data would fit
better with lower tensor analyzing powers, such as those published
by Becker. We could not use Becker's tensor analyzing powers because
our measurements was performed at different
energies.\\

Although $P_Z$ depends on the value of $P_{ZZ}$ ($\epsilon_1$ in
Eq.\,(\ref{eq:spin1asym})), this dependence is very weak. Having the
wrong $P_{ZZ}$ changes $P_Z$ by a negligible amount.\\

The vector polarization that the LDP measures is reliable because
the vector analyzing power is energy independent, matches the
theoretical prediction, and does not depend on our choice of model
for effective analyzing powers. \\

Finally, the LDP measures $P_Z$'s of vector polarized beams that are
on average 7\% lower than the LSP. This relative difference can
originate from unknown systematic uncertainties of the LSP or LDP,
or because of polarization change in space (between the
polarimeters) or in time (between the measurements).\\

\newpage
\section{Conclusions}
We built a Low Energy Deuteron Polarimeter (LDP), based on a
reaction that received little attention for polarimetry. We find
that the LDP is well suited to measure asymmetries but presently
lacks a proper tensor analyzing power calibration.\\

Data from our polarization cross-check experiment cannot answer
whether there are polarization changes during acceleration,
because our IBP data are, at this point, preliminary.\\

The relative difference, of vector polarized beams, between the LSP
and LDP was 7\% in our experiment.

\subsection{Future Improvements}

We can point out future improvements to the experiment we performed,
and to future versions of the LDP.

\subsubsection{Polarimeter Cross-Check Experiment}
\begin{itemize}
\item The online analysis program of the LSP, although fast, often
returns inaccurate polarization and uncertainties in polarization.
Offline analysis, which was done here, decreased the uncertainties
(sometimes by a factor of 10) and changed the polarization (mostly
the tensor component of polarization). The online analysis program
could be made more accurate.

\item Use all the polarization states that POLIS can provide (13),
not just pure vector (2) and tensor beams (2) but mixed
vector-tensor beams (9), for additional systematic checks between
polarimeters.
 \item Our conclusions rely to a large extent on the fact that the
 beam polarization from the source was constant during the experiment. During next
 experiment the polarization should be monitored with the LSP
 before and after each run.
 \end{itemize}

\subsubsection{LDP}
\begin{itemize}

\item Find better tensor analyzing powers. Either by
improving the model that calculates effective analyzing power, or by
using a better data set of published tensor analyzing powers.

\item Measure the energy dependence of the vector analyzing power.
We expect it to be energy independent, but do not know to what
level.

 \end{itemize}

\newpage
\section{Acknowledgement}
I would like to thank a number of people. Starting with Johan
Messchendorp, my supervisor, who was involved in every aspect of
this project. His experience and insight were extremely valuable to
me. Nasser Kalantar-Nayestanaki, my professor, for his stimulating
and energetic character. Hossein Mardanpour, my colleague, for
constantly helping me with computers. Rob Kremers and Hans Beijers
for useful discussions. The Few-Body Physics Group for their
conviviality. Finally my parents for their support and patience.

\end{document}